\documentstyle[psfig]{l-aa}

\def\Msun{M$\sb{\odot}$}

\def\Mbol{$M\sb{\rm bol}$}
\def\Mv{$M\sb{\rm v}$}

\def\Teff{$T\sb{\rm eff}$}
\def\Hp{$Hp$}
\def\VI{$(V-I)_{\rm C}$}

\input epsf

\begin{document}

\title{The HIPPARCOS Hertzsprung-Russell diagram of S stars: 
Probing nucleosynthesis and dredge-up\thanks{Based on data from the HIPPARCOS
astrometry satellite}}
\author{S.~Van Eck\inst{1}
\and
A.~Jorissen\inst{1}
\and
        S.~Udry\inst{2}
\and    
        M.~Mayor\inst{2}
\and    
        B.~Pernier\inst{2}
}
\offprints{S.~Van Eck (svaneck@astro.ulb.ac.be)}
\institute{
Institut d'Astronomie et d'Astrophysique,
                  Universit\'e Libre de Bruxelles, C.P. 226,
                  Boulevard du Triomphe,
                  B-1050 Bruxelles,
                  Belgium
\and 
Observatoire de Gen\`eve, 51 Chemin des Maillettes, CH-1290 Sauverny, Switzerland
}

\date{Received date; accepted date}

\thesaurus{06( 08.16.4 -- 08.08.1 -- 08.05.3 -- 08.06.3 -- 08.12.3) }

\maketitle
\markboth{S. Van Eck et al.: The HR diagram of S stars}
{S. Van Eck et al.: The HR diagram of S stars}

\begin{abstract}
HIPPARCOS trigonometrical parallaxes make it possible to compare the location of
Tc-rich and Tc-poor S stars in the Hertzsprung-Russell (HR) diagram: 
Tc-rich S stars are found to be cooler and intrinsically brighter than Tc-poor S
stars.

The comparison with the Geneva evolutionary tracks reveals that the line marking
the onset of thermal pulses on the asymptotic giant branch (AGB) matches well
the observed limit between Tc-poor and Tc-rich S stars.
Tc-rich S stars are, as expected, identified with  thermally-pulsing AGB stars of
low and intermediate masses, whereas Tc-poor S stars comprise mostly low-mass
stars (with the exception of 57 Peg) located either on the red giant branch or on
the early AGB. Like barium stars, Tc-poor S stars are known to belong exclusively
to binary systems, and their location in the HR diagram is consistent with the
average mass of $1.6\pm0.2$ \Msun\ derived from their orbital mass-function distribution
(Jorissen et al. 1997, A\&A, submitted). 

A comparison with the S stars identified in the Magellanic Clouds
and in the Fornax dwarf elliptical galaxy reveals that
they have luminosities similar to the galactic Tc-rich S stars.
However, most of the surveys of S stars in the external systems did
not reach the lower luminosities at which galactic Tc-poor S stars are
found. The deep Westerlund survey of carbon stars in the SMC uncovered
a family of faint carbon stars that may be the analogues of the 
low-luminosity, galactic Tc-poor S stars.

\end{abstract}


\section{ Introduction }

The S stars are late-type giants whose spectra resemble those of M 
giants, with the addition of distinctive molecular bands of ZrO 
(Merrill 1922). Detailed abundance analyses of S stars (e.g., Smith \& Lambert
1990) have shown that the overabundance pattern for the elements
heavier than Fe bears the signature of the s-process 
nucleosynthesis (see K\"appeler et al. 1989). 
Furthermore, the C/O
ratio in S stars is intermediate between that  of normal M giants
($\sim 0.2$) and that of carbon stars ($> 1$).  
S stars were therefore traditionally considered as
transition objects between normal M giants and carbon stars on the
asymptotic giant branch (AGB) (e.g., Iben \& Renzini 1983). Internal
nucleosynthesis  processes related to the thermal pulses (a recurrent thermal
instability affecting the He-burning shell in AGB stars)  could
possibly produce the s-process elements (see the discussion by
Sackmann \& Boothroyd 1991), and the `third dredge-up'
(the descent of the lower boundary of  the convective envelope into
the region formerly processed by the thermal pulse) 
could bring them to the surface.  
The simple M--S--C evolution sequence faces, however, several problems, as
discussed by   
e.g. Lloyd Evans (1984), Mould \& Aaronson (1986) and Willems \& de
Jong (1986).
The observation of Tc lines (an element with no stable isotopes) in
some but not all S stars (Merrill 1952; Little et al. 1987) 
raised another major problem. If the s-process really occurred during
recent thermal pulses in S stars, Mathews et
al. (1986) predicted that Tc should be observed at the surface
along with the other s-process elements.

A breakthrough in our understanding of the evolutionary status of S
stars came with the realization that Tc-poor S stars are
all members of binary systems (Brown et al. 1990; Jorissen et
al. 1993; Johnson et al. 1993), and are likely
the cooler analogues of the barium stars, a family of peculiar G--K giants first
identified by Bidelman \& Keenan (1951). 
Barium stars and S stars share the same
abundance peculiarities (e.g., McClure 1984) and high rate of binaries
with suspected white dwarf (WD) companions (B\"{o}hm-Vitense et al.
1984; McClure \& Woodsworth 1990; Jorissen et al. 1997a).  
Both families are believed to owe their chemical peculiarities
to the transfer of s-process-rich matter from the former AGB star (the progenitor
of the current WD) to the (main sequence) progenitor of the current 
chemically-peculiar red giant.

It is therefore currently believed that
two very different kinds of stars are found among S stars: Tc-rich {\it
intrinsic} S stars that owe their chemical peculiarities to internal
nucleosynthesis processes occurring during thermal pulses on the AGB, and 
Tc-poor {\it extrinsic} S
stars, all members of binary systems with mass transfer 
responsible for the pollution of  the
S star envelope. Contrarily to intrinsic S stars that need to be thermally-pulsing
AGB (TP-AGB) stars, extrinsic S stars
may populate the first-ascent red giant branch (RGB), prior to the core He
flash, or the early AGB (E-AGB) preceding the TP-AGB, since the mass-transfer
scenario does not set
any constraint on the {\it current} evolutionary stage of the extrinsic S star.
A direct check of these predictions has been
hampered till now by the 
difficulty in evaluating the absolute magnitude of S stars.
Methods used so far include individual parallaxes ($\chi$
Cyg: Stein 1991), membership in a binary system with a detected main
sequence companion ($\pi^1$ Gru: Feast 1953; 57 Peg: Hackos \& Peery
1968; T Sgr: Culver \& Ianna 1975), membership in a moving group
($\pi^1$ Gru, R Hya: Eggen 1972a; R And, HR 363, $o^1$ Ori: Eggen
1972b), membership in a cluster or association (WY Cas: Mavridis 1960;
TT9, TT12: Feast et al. 1976), CaII K-line emission width (HD 191630,
57 Peg: Warner 1965; $o^1$ Ori: Boesgaard 1969). 
Most of these individual estimates were used by Scalo (1976) to locate S stars
in the Hertzsprung-Russell (HR) diagram, with the conclusion that
they fall above the luminosity threshold for TP-AGB
stars, as derived from the early AGB model calculations available to Scalo.
Although Scalo's conclusion appears to support the customary M-S-C sequence,
statistical estimates 
based on the kinematics or space distribution of S stars pointed out a systematic
difference between the average luminosity of non- or weakly-variable S stars and
of Mira S stars, the latter being about 3 visual magnitudes brighter than 
the former (Takayanagi 1960; Yorka \& Wing 1979). Such a segregation apparent in
these early studies might already hint at the current dichotomy between extrinsic
and intrinsic S stars.

The rather heterogeneous set of former luminosity determinations for
S stars, along with our revised understanding of their evolutionary status,
prompted
the present study, based on the trigonometrical parallaxes provided by the
HIPPARCOS satellite.
Its aim is to locate both kinds of S stars in the HR diagram, 
and to compare their respective locations with stellar evolutionary tracks.

Our HIPPARCOS sample of S stars is described in Sect.~\ref{Sect:HIP}.
The method used for deriving their bolometric magnitudes is 
detailed in Sect.~\ref{Sect:colors}.  The HR diagram of S stars is discussed  in
Sect.~\ref{section:HR}, with special emphasis on the comparison 
with theoretical evolutionary
tracks and with S and C stars in the Magellanic Clouds. Finally,
the correlation between the infrared colours of S stars and their location on the
giant branches is discussed  in Sect.~\ref{Sect:IR}.

\section{S stars in the HIPPARCOS catalogue}
\label{Sect:HIP}

A full discussion of the HIPPARCOS mission can be found in the HIPPARCOS
catalogue (ESA 1997). Cross-identifications between the HIPPARCOS catalogue
and the {\it General Catalogue of Galactic S Stars} (Stephenson 1984; GCGSS)
yield 63 stars in common between the two catalogues. Of these, 22 are part of
the HIPPARCOS general survey (a systematic monitoring of all stars down to
magnitudes 7.3 -- 9, depending on galactic latitude and spectral type), and 41
belong to samples included in the HIPPARCOS Input Catalogue (Turon et al. 1992abc) 
for a particular purpose. The sample of S stars studied
in this paper can thus in no way be considered as a complete sample, 
since it rather reflects
the particular interests prevailing at the time of construction of the HIPPARCOS Input
Catalogue. A comparison with the GCGSS indicates that the HIPPARCOS
sample of S stars is nevertheless complete down to $V = 7.5$. 

Table~\ref{Tab1:sample} lists various identifications of the S stars considered
in this paper. The different columns contain the following data:\\
1. HIPPARCOS Input Catalogue (HIC) number;\\
2-3. GCGSS and HD numbers;\\
4. variable name in the the {\it General Catalogue of Variable Stars} (Kholopov
et al. 1985;  GCVS); \\
5. variability type from the GCVS;\\
6. spectral type from the GCGSS;\\
7. presence (y) or absence (n) of technetium lines (from Little et al. 1987,
Smith \& Lambert 1988, from the compilation of Jorissen et al. 1993, or from
Van Eck \& Jorissen (in preparation) in the case of HIC 28297);\\
8-9. annual parallax $\pi$ and its standard error $\sigma_\pi$ (both expressed in
milliarcseconds).
 
Table~\ref{Tab2:sample} lists the basic
photometric properties of the S stars in the HIPPARCOS sample. The various columns
contain the following data:\\ 
1. HIC number;\\
2-3. median \Hp\ magnitude and its standard error $\sigma_{\rm Hp}$;\\
4. colour excess $E_{\rm B-V}$;\\
5. Johnson $V_{\rm J}$ magnitude from the HIPPARCOS catalogue;\\
6. dereddened $(V-K)_0$ colour index;\\
7. dereddened $K_0$ magnitude;\\
8. bolometric correction $BC_{\rm K}$ to the $K$ magnitude;\\
9. bolometric magnitude \Mbol\ (see Sect.~\ref{Sect:colors} for details    
about the derivation of the quantities listed in columns 4--9);\\
10. method used for deriving the bolometric correction: (1) by integrating under
the energy curve, 
or (2) from the $(BC_{\rm K}, V-K)$ relation of Bessell \& Wood (1984);\\
11. source for the HIPPARCOS $(V-I)_{\rm C}$ colour 
[A-G: from $VRI_{\rm C}$ photometry; H-K: from $UBV_{\rm J}$ photometry;
L-P: from HIPPARCOS and Star Mapper photometry; Q: specific treatment
applied to long-period variables;
R: from spectral type; see the HIPPARCOS and Tycho Catalogues 
(ESA 1997, Vol. 1, {\it Introduction and Guide to the Data}) for details];\\
12. duplicity flag (field H59 of the HIPPARCOS catalogue; 
see Sect.~\ref{Sect:errors});\\
13. other identifications ('Hen' refers to the survey of S stars by
Henize 1960).
          
\renewcommand{\baselinestretch}{0.7}
\begin{table*}
\caption[]{\label{Tab1:sample}
S stars in the HIPPARCOS catalogue: Identifications and parallaxes
}
\small{
\begin{tabular}{rrrllllrr}
   HIC & GCGSS & HD & Var & Var type & Sp. & Tc & $\pi$ & $\sigma_\pi$ \cr\\
\hline\\
   621&3    &310   &        &     &S3,1       &n &  2.50& 0.69  \cr  
  1728&8    &1760  &T Cet   &SRc  &M5-6Se; M5-6Ib-II&y &  4.21& 0.84  \cr  
  1901&9    &1967  &R And   &M    &S5-7/4-5e  &y & -0.06& 6.49   \cr  
  5091&22   &6409  &        &     &M2wkS      &n &  2.43& 0.89   \cr  
  5772&26   &7351  &        &     &S3/2       &n &  3.21& 0.82   \cr  
  8876&45   &      &        &     &S3/1       &n & -1.92& 1.50   \cr  
 10687&49   &14028 &W And   &M    &S7/1e      &y & -1.17& 3.17   \cr  
 17296&79   &22649 &BD Cam  &Lb   &S4/2       &n &  6.27& 0.63   \cr  
 21688&104  &29704 &        &     &S:         &n &  1.84& 1.00   \cr  
 22667&114  &30959 &$o^1$ Ori&SRb &S3/1       &y &  6.02& 0.94   \cr  
 25092&133  &35155 &        &     &S4,1       &n &  1.32& 0.99   \cr  
 26718&149  &37536 &NO Aur  &Lc   &M2S        &y &  2.38& 0.97   \cr  
 28297&178  &40706 &        &     &S2,1       &n &  1.17& 0.88   \cr  
 30301&212  &44544 &FU Mon  &SR   &S7/7 (SC)  &  &  0.30& 1.58   \cr  
 32627&260  &49368 &V613 Mon&SRb: &S3/2       &n &  1.65& 1.11   \cr  
 32671&265  &49683 &        &     &M4S        &  & -0.18& 1.07   \cr  
 33824&283  &51610 &R Lyn   &M    &S5/5e      &  & -3.39& 1.79   \cr  
 34356&307  &53791 &R Gem   &M    &S4*1       &y & -6.22& 6.50   \cr  
 35045&312  &54587 &AA Cam  &Lb   &M5S        &y &  1.24& 1.02   \cr  
 36288&347  &58521 &Y Lyn   &SRc  &M6S        &y &  4.03& 1.33   \cr  
 37521&382  &61913 &NZ Gem  &SR   &M3S        &n?&  3.19& 0.79   \cr  
 38217&411  &63733 &        &     &S4/3       &y?&  0.00& 0.99   \cr  
 38502&422  &64332 &NQ Pup  &Lb   &S5/2       &y &  3.01& 1.11   \cr  
 38772&436  &      &SU Pup  &M    &S4,2       &  & -1.75& 1.81   \cr  
 40977&494  &70276 &V Cnc   &M    &S3/6e      &  & 26.58&42.74   \cr  
 45058&589  &78712 &RS Cnc  &SRc: &M6S        &y &  8.21& 0.98   \cr  
 54396&722  &96360 &        &     &M3         &n &  2.10& 0.90   \cr  
 59844&788  &      &BH Cru  &M    &S5,8e (SC) &  &  1.64& 0.99   \cr  
 62126&803  &110813&S UMa   &M    &S3/6e      &y &  0.63& 0.94   \cr  
 64613&815  &      &        &     &S3,3       &  & -1.90& 2.99   \cr  
 64778&816  &115236&UY Cen  &SR   &S6/8 (CS)  &  &  1.66& 1.04   \cr  
 66783&826  &118685&        &     &S6,2       &n &  3.42& 0.65   \cr  
 67070&829  &      &        &     &M1wkS      &n &  2.27& 1.06   \cr  
 68837&     &      &U Cir   &SR   &C          &  & -1.92& 1.51   \cr  
 71348&804  &110994&BQ Oct  &Lb:  &S5,1       &  &  2.08& 0.57   \cr  
 72989&867  &131217&        &     &S6,2       &n &  4.24& 1.02   \cr  
 77619&903  &142143&ST Her  &SRb  &M6.5S      &y &  3.22& 0.75   \cr  
 81970&937  &      &        &     &M2S        &n &  3.32& 1.14   \cr  
 82038&938  &151011&        &     &Swk        &n &  4.30& 1.07   \cr  
 87850&     &163990&OP Her  &SRb  &M6S        &y &  3.26& 0.54   \cr  
 88620&1014 &164392&        &     &           &  &  2.09& 1.06   \cr  
 88940&1023 &165774&        &     &S4,6       &n &  0.23& 1.50   \cr  
 89316&1025 &165843&        &     &S2,1       &  &  1.47& 1.05   \cr  
 90723&1053 &170970&        &     &S3/1       &y &  1.83& 0.67   \cr  
 94706&1117 &180196&T Sgr   &M    &S5/6e      &y &-31.67& 9.28   \cr  
 97629&1165 &187796&$\chi$ Cyg &M &S7/1.5e    &y &  9.43& 1.36   \cr  
 98856&1188 &190629&AA Cyg  &SRb  &S6/3       &y &  0.86& 0.88   \cr  
 99124&1192 &191226&        &     &M1S-M3SIIIa&n &  0.39& 0.71   \cr  
 99312&1194 &191589&        &     &S:         &n &  2.25& 0.77   \cr  
 99758&1195 &191630&        &     &S4,4       &y &  1.18& 0.81   \cr  
100599&1211 &      &V865 Aql&M    &S7,2       &  & -1.28& 1.91   \cr  
101270&1224 &195665&AD Cyg  &Lb   &S5/5       &  & -0.96& 1.15   \cr  
103476&1254 &199799&        &     &MS         &  &  2.16& 0.82   \cr  
110146&1292 &211610&X Aqr   &M    &S6,3e:     &  & -4.01& 5.73   \cr  
110478&1294 &212087&$\pi^1$ Gru&SRb&S5,7:     &y &  6.54& 1.01   \cr  
112227&1304 &215336&        &     &Swk        &n &  0.74& 0.92   \cr  
112784&1309 &      &SX Peg  &M    &S3/6e      &  &  2.12& 2.99   \cr  
113131&1315 &216672&HR Peg  &SRb  &S4/1       &y &  3.37& 0.94   \cr  
114347&1322 &218634&57 Peg  &SRa  &M4S        &n &  4.28& 0.88   \cr  
115965&1334 &      &        &     &S2/3:      &n &  1.72& 1.26   \cr\\
\noalign{HIPPARCOS close visual binaries}\\
 19853&89   &26816 &        &     &S          &y &  3.79& 1.05   \cr  
 27066&157  &      &        &     &S          &  &  1.13& 3.50   \cr\\  
\noalign{Misclassified S star}\\
 42650&544  &      &        &     &MS$^a$     &  & 32.13& 1.50   \cr\\  
\noalign{Comparison barium star}\\
 68023&     &121447&        &     &K7IIIBa5;
S0&n$^b$&\multicolumn{2}{c@{\mbox{}}}{428$\pm$71 pc $^c$}\cr\\
\hline
\end {tabular}
}
Remarks:\\
a: HIC 42650 is an early M dwarf rather than a MS star\\
b: Little et al. (1987)\\
c: Distance derived by Mennessier et al. (1997) from a maximum-likelihood
estimator based on the HIPPARCOS parallax
\end{table*}

\begin{table*}
\caption[]{\label{Tab2:sample}
S stars in the HIPPARCOS catalogue: Photometry
}
\small{
\begin{tabular}{rrllrrrllllll}
   HIC & $Hp$ & $\sigma_{\rm Hp}$ &  $E_{\rm B-V}$ & $V_{\rm J}$ &
$(V-K)_0$ & $K_0$ & $BC_{\rm K}$ &\multicolumn{2}{c}{\Mbol} & $S_{\rm V-I}$ & Dupl
& Other ident. \cr\\
\hline\\
   621& 7.570&0.003&0.000& 7.50&4.38 &3.12 &2.74&-2.14&   2   &O  &   & Hen 1\cr
  1728& 5.439&0.033&0.000& 5.61&6.5  &-0.89&2.91&-4.85&   1   &O  &   & \cr
  1901&10.705&0.011&     &10.71&     &-0.11&3.49&     &   2   &K  &V  &         
                     \cr  
  5091& 7.462&0.004&0.042& 7.44&5.04 &2.27 &2.84&-2.95&   2   &O  &   &\cr
  5772& 6.395&0.004&0.049& 6.33&4.65 &1.53 &2.68&-3.25&   1   &F  &   &HR 363\cr
  8876& 9.072&0.002&     & 8.97&     &     &    &     &   2   &L  &  
&+21$^\circ$255                        \cr  
 10687& 8.075&0.226&     & 8.60&     &0.7  &3.49&     &   2   &O  &   &\cr
 17296& 5.095&0.003&0.041& 5.06&4.78 &0.15 &2.77&-3.09&   1   &C  &O  &HR 1105\cr
 21688& 8.225&0.009&0.000& 8.23&5.64 &2.59 &2.91&-3.16&   2   &O  &   &Hen 3    
                   \cr  
 22667& 4.702&0.005&0.121& 4.71&4.96 &-0.62&2.85&-3.87&   1   &O  &V  &\cr
 25092& 6.882&0.011&0.128& 6.82&4.46 &1.97 &2.76&-4.66&   2   &F  &   &\cr
 26718& 6.243&0.011&0.245& 6.23&4.67 &0.8  &2.74&-4.57&   1   &C  &   &\cr
 28297& 8.987&0.002&0.029& 8.88&     &     &    &     &   2   &L  &   &Hen 7\cr
 30301& 8.520&0.032&0.327& 9.76&7.35 &1.4  &3.03&     &   1   &O  &   &\cr
 32627& 7.749&0.004&0.051& 7.73&4.93 &2.64 &2.83&-3.43&   2   &O  &   &\cr
 32671& 8.232&0.006&     & 8.38&     &2.07 &3.49&     &   2   &O  &   &\cr
 33824& 9.922&0.029&     & 9.93&     &1.91 &3.49&     &   2   &K  &   &\cr
 34356& 7.529&0.456&     & 7.53&     &2.12 &3.49&     &   2   &P  &V  &\cr
 35045& 7.578&0.007&0.039& 7.69&6.01 &1.56 &2.95&-5.01&   2   &O  &V  &\cr
 36288& 6.897&0.005&0.058& 7.27&7.63 &-0.54&3.08&-4.43&   1   &O  &V  &\cr
 37521& 5.587&0.003&0.027& 5.55&4.85 &0.61 &2.86&-4.01&   1   &G  &   &HR 2967\cr
 38217& 7.981&0.002&     & 7.90&     &     &3.49&     &   2   &F  &   &\cr
 38502& 7.532&0.015&0.072& 7.52&5.17 &2.13 &2.86&-2.61&   2   &F  &V  &\cr
 38772& 9.609&0.218&     & 9.64&     &2.64 &3.49&     &   2   &K  &   &Hen 32   
                   \cr  
 40977& 9.262&0.152&0.004& 9.25&     &     &3.49&     &   2   &K  &X  &\cr
 45058& 5.450&0.026&0.002& 6.04&7.74 &-1.71&3.07&-4.06&    1  &O  &   &\cr
 54396& 8.075&0.006&0.000& 8.05&     &     &    &     &    2  &O  &   &\cr
 59844& 7.737&0.118&0.163& 7.74&     &     &3.49&     &    2  &K  &   &Hen 120  
        \cr  
 62126& 8.909&0.148&0.000& 8.94&5.95 &2.99 &2.87&     &    1  &K  &   &\cr
 64613&11.404&0.021&     &11.33&     &2.85 &3.49&     &    1  &K  &   & Hen 134, 
-30$^\circ$10427           \cr  
 64778& 6.787&0.033&0.088& 6.85&6.09 &0.49 &2.85&-5.55&    2  &L  &V  &Hen 135\cr
 66783& 6.832&0.007&0.062& 6.91&5.82 &0.9  &2.93&-3.49&    2  &O  &   & Hen 138\cr
 67070& 8.541&0.002&0.009& 8.43&     &     &    &     &    2  &L  &  
&-2$^\circ$3726                        \cr  
 68837& 9.609&0.013&     & 9.54&     &     &    &     &    2  &J  &V  &C* 2142\cr
 71348& 6.806&0.005&0.106& 6.82&5.16 &1.33 &2.86&-4.21&    2  &O  &   &Hen 127\cr
 72989& 7.425&0.004&0.204& 7.45&     &     &    &     &    2  &O  &   &Hen 150\cr
 77619& 6.920&0.026&0.000& 7.69&8.55 &-0.86&3.11&-5.20&    2  &O  &   &\cr
 81970& 7.803&0.004&0.173& 7.99&5.63 &1.83 &2.91&-2.64&    2  &O  &  
&$-13^\circ$4495                       \cr  
 82038& 6.689&0.002&0.190& 6.60&4.59 &1.43 &2.78&-2.62&    2  &L  &   &\cr
 87850& 6.105&0.010&0.022& 6.22&6.09 &0.06 &2.92&-4.45&    1  &O  &   &\cr
 88620& 8.448&0.004&0.072& 8.39&     &     &    &     &    2  &O  &   &Hen 183\cr
 88940& 8.211&0.003&0.186& 8.17&     &     &    &     &    2  &O  &   &Hen 186\cr
 89316& 8.412&0.005&0.082& 8.37&     &     &    &     &    2  &O  &   &Hen 187\cr
 90723& 7.457&0.002&0.041& 7.42&4.88 &2.41 &2.82&-3.45&    2  &O  &   &\cr
 94706&10.826&0.223&     &10.78&     &1.05 &3.14&     &    2  &K  &V  &\cr
 97629& 6.169&0.168&0.004& 7.91&7.69 &-1.73&3.27&-3.58&    1  &O  &V  &\cr
 98856& 8.159&0.037&0.491& 8.16&6.26 &0.39 &2.94&     &    2  &K  &   &\cr
 99124& 7.374&0.002&0.491& 7.28&3.4  &2.37 &2.52&     &    2  &L  &   &\cr
 99312& 7.368&0.001&0.081& 7.26&     &     &    &     &    2  &H  &   &\cr
 99758& 6.757&0.004&0.029& 6.74&5.05 &1.6  &2.71&-5.33&    2  &O  &   &Hen 197  
                   \cr  
100599& 9.642&0.134&     &10.34&     &1.49 &3.49&     &    2  &Q  &  
&+0$^\circ$4492                        \cr  
101270& 8.607&0.011&     & 8.61&     &1.18 &3.49&     &    2  &K  &   &\cr
103476& 7.248&0.016&0.071& 7.36&5.99 &1.15 &2.95&-4.22&    2  &O  &   &\cr
110146&10.188&0.179&     &10.82&     &2.9  &3.49&     &    2  &Q  &V  &         
                     \cr  
110478& 5.495&0.011&0.000& 6.42&8.55 &-2.13&3.12&-4.93&    1  &O  &V  &Hen 202\cr
112227& 7.929&0.001&0.089& 7.82&     &     &    &     &    2  &L  &   &         
                     \cr  
112784& 9.241&0.176&0.056& 9.25&     &     &    &     &    2  &K  &V  &         
                     \cr  
113131& 6.346&0.016&0.050& 6.39&5.47 &0.77 &2.86&-3.73&    1  &O  &V  &HR 8714\cr
114347& 5.033&0.011&0.024& 5.05&5.47 &-0.5 &3   &-4.34&    1  &F  &   &\cr
115965& 9.536&0.003&0.056& 9.43&     &     &    &     &    2  &K  &  
&+28$^\circ$4592                       \cr\\
\noalign{HIPPARCOS close visual binaries}\\
 19853& 7.607&0.010&0.120& 7.90&     &     &    &     &   2   &O  &C  &
+23$^\circ 654$\cr
 27066&11.274&0.010&0.327&11.12&     &     &    &     &    2  &R  &C  &         
                     \cr\\
\noalign{Misclassified S star}\\                  
 42650&11.118&0.005&0.009&11.03&     &     &    &     &    2  &I  &   &         
                     \cr\\
\noalign{Comparison barium star}\\                
 68023&      &     &0.056&     &3.78 &4.090&2.69&-1.40&    1  &   &   &\cr\\
\hline
\end {tabular}
}
\end{table*}
\renewcommand{\baselinestretch}{1} 

\begin{figure}
   \begin{center}
   \leavevmode
   \centerline{\psfig{file=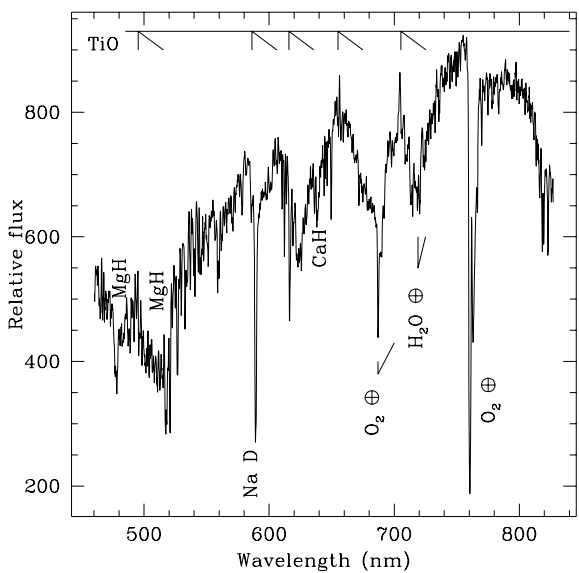,width=8.0cm,height=8.0cm}}
   \end{center}

\caption[]{\label{Fig:HIC42650}
The spectrum of HIC 42650 = GCGSS 544, with the principal spectral features
identified
}
\end{figure}

One star (HIC 42650 = GCGSS 544) appears outlying in many respects. With the
largest parallax in the sample and the second faintest \Hp\ magnitude, 
it is intrinsically much fainter than all the
other stars in the sample. 
It is also somewhat bluer, with $(B-V)_J=1.39$ and $(V-I)_C=1.65$, compared to  
average colour indices of $\langle (B-V)_J\rangle =1.79$ and
$\langle (V-I)_C\rangle =2.72$ for the whole sample.
A low-resolution spectrum ($\Delta \lambda \sim 0.3$ nm) of this star, covering
the spectral range 440 -- 820 nm, has been obtained at the {\it European Southern
Observatory} (ESO, La Silla, Chile) on the 1.52-m telescope equipped
with the Boller \& Chivens spectrograph (grating \#23 + filter GG 420; 114 \AA\
mm$^{-1}$) and a Loral/Lesser thinned, UV-flooded 2048x2048 CCD (CCD\#39; 15
$\mu$m pixels). The flux response curve of the system has been calibrated using
the spectrophotometric standard star LTT 4816.

The spectrum of GCGSS 544 appears to be that of a M0-M1 dwarf, as can be seen on
Fig.~\ref{Fig:HIC42650} from the strong MgH ($\lambda 478.0$ and $\lambda 521.1$)
and CaH ($\lambda 638.2$ and $\lambda 638.9$) bands, as well as from the strong
Na D line superimposed on moderately strong TiO bands (Jaschek \& Jaschek 1987).
This classification is consistent with its absolute
visual magnitude \Mv $= 8.6$ derived from its HIPPARCOS parallax and $V_J$
magnitude. That star is thus misclassified as
S star, since  there is no trace whatsoever of ZrO bands in its spectrum.

The case of T Cet (= HIC 1728) is also conflicting, since it was
classified as M5-6Se in the original paper by Keenan (1954) defining the
S class, and reclassified as M5-6Ib-II in the Michigan Spectral Survey
(Houk \& Cowley 1975). The spectra of supergiants and weak S-type stars look very
similar at the low plate dispersions 
used in classification work, and are therefore easily confused (e.g.,
Lloyd Evans
\& Catchpole 1989). 
T Cet has nevertheless been kept in our final list until higher resolution 
spectra resolve these equivocal classifications.

The star HIC 27066 (= GCGSS 157) has HIPPARCOS colours that do not match those of
an S star [$(B-V)_J=0.80$ and $(V-I)_C=0.83$]. HIPPARCOS found it to be a close
visual binary with a separation of 0.262 arcsec, the companion being 1.55 mag
fainter in the \Hp\ band. The colour indices measured by HIPPARCOS
correspond to the composite light from the system.
 
The only other S star found by HIPPARCOS to be a close visual binary
(with an angular separation of 0.18 arcsec) is HIC 19853. 
Although its colour indices are not atypical for a late-type star,
the measured colours must clearly be composite since the companion is
only 0.35 mag fainter in the \Hp\ band.

These two stars have been excluded from our final sample.
Among the 61 remaining S stars, 20 do exhibit technetium lines
in their spectrum, 21 do not, and the Tc content is unknown
for the remaining 20 stars.

The only trigonometrical parallax for an S star with a small relative error
available in the literature prior to the HIPPARCOS mission is that obtained by
Stein (1991) for the Mira S star  $\chi$ Cyg. His value ($8.8\pm 1.9$ mas) is in
good agreement with the HIPPARCOS parallax listed in Table~\ref{Tab1:sample}.

\section{Colours and bolometric magnitudes}
\label{Sect:colors}

\subsection{Colours}

Although the $(B-V)_{\rm J}$  and $(V-I)_{\rm C}$ (where the
subscripts J and C stand for the Johnson and Cousins photometric
systems, respectively) 
colour indices are directly provided by the
HIPPARCOS catalogue, we felt that the $(V-K)_{\rm J}$ 
colour index, derived on a case-by-case
basis, is a more appropriate temperature indicator.   
For most of the stars in our sample, the $(V-I)_{\rm C}$ index listed in the
HIPPARCOS catalogue has been derived from a fiducial relation \VI $=f(B-V)_{\rm
J}$
calibrated on normal M giants (as indicated by the flags H-P in column $S_{\rm V-
I}$ of Table~\ref{Tab2:sample}). Several S stars in our sample, however, fall
outside the validity range of such a calibration, so that \VI\ as provided by the
HIPPARCOS catalogue is not a reliable temperature indicator for our purpose.

As discussed by Ridgway et al. (1980), $B-V$ for late-type stars is not a good
temperature indicator either,
because the temperature has opposite effects on $B-V$: when the temperature decreases,
the absorption in the $V$ band due to TiO bands increases (thus decreasing $B-V$),
whereas the black body continuum tends to increase the $B-V$ index.
By contrast, these two effects act together to increase $V-K$. 

The $V-K$ index has therefore been constructed from 
individual $K$ magnitudes collected from the literature, mostly from the 
{\it Two-Micron Sky Survey} (Neugebauer \& Leighton 1969; TMSS), 
but also from Wing \& Yorka (1977), Catchpole et al. (1979),
Chen et al. (1988), Noguchi et al. (1991)  and
Fouqu\'e et al. (1992); it is available for 46 stars as listed in
Table~\ref{Tab2:sample}.
The adopted $V$ magnitude corresponds to the $V_{\rm J}$ magnitude listed in the
HIPPARCOS catalogue. 
It has been obtained by the HIPPARCOS reduction consortia 
from a $(H_{\rm p}-V_{\rm J}$, \VI) 
relation, with \VI\ derived by different methods as listed in column $S_{\rm V-I}$
of Table~\ref{Tab2:sample}.  
According to the HIPPARCOS reduction consortia (ESA 1997, Vol. 1, {\it
Introduction and
Guide to the Data}), 
that relation is quite well defined 
from classical photometry in
the range $-0.4 <$  \VI $< 3.0$ (which holds for 47 S stars),
with uncertainties of less than 0.01 mag.
The red extension down to \VI $=5.4$  (15 S stars)
was defined using observations of Mira variables devoted to that purpose,
resulting in an uncertainty of 0.03--0.05 mag. 
For indices down to \VI $=9.0$ [only $\chi$ Cyg is concerned,
with \VI $=6.1$], the HIPPARCOS reduction consortia adopted a linear
extrapolation of the previous relation. To avoid the uncertainties inherent to
such an extrapolation, a (flux) average of the numerous $V_{\rm J}$ measurements
of $\chi$ Cyg available in the literature has been preferred over the linear
extrapolation. 

In summary, the uncertainty on $V_{\rm J}$ introduced by the colour
transformation applied on \Hp\ is largely offset by the fact that
the \Hp\ magnitude is a good time average over a uniform 
period of time (the duration of the HIPPARCOS  mission), the same for all the
stars. As discussed in Sect.~\ref{Sect:errors}, the intrinsic
variability of S stars is indeed a major source of uncertainty on their
location in the HR diagram.

\subsection{Bolometric magnitudes}
\label{Sect:BC}

Since most of the flux from S stars is radiated in the near-infrared, 
the bolometric correction is best determined from near-infrared magnitudes. 
Moreover, the near-infrared and bolometric variability  
is much smaller than the visual amplitude of variations
(Mira-type variables which vary typically by 6 to 8 mag 
in the visual region, vary by less than 1 mag in the
$K$ band; Feast et al. 1982).
Therefore, the bolometric correction $BC_{\rm K}$ in the $K$ band (defined
as \Mbol\ $= K + BC_{\rm K}$) has been
adopted in this work (and
listed in Table~\ref{Tab2:sample}).

In order to compute bolometric corrections, 
an extensive set of magnitudes ranging from the ultraviolet to the far IR has
been collected from the literature. 
This set includes Johnson $UBVRIJHKLM$ magnitudes when available, as well as 
good quality fluxes at
4.2, 11.0, 19.8 and 27.4~$\mu$m from the  
Revised Air Force Four-Color Infrared Sky Survey
(Price \& Murdock 1983), and
four-colour infrared photometry from Gillett \& Merrill (1971).
The IRAS 12, 25, 60 and 100 $\mu$m fluxes (with a quality flag 3) 
from the second edition
of the {\it Point Source Catalogue} (IRAS Science Team 1988) were also used, or
when available,
the reprocessed IRAS fluxes provided by Jorissen \& Knapp (1997).
When several measurements in the same filter are available, a flux average has
been computed.

All magnitudes bluer than 4.8$\mu$m have been corrected for
interstellar reddening and absorption,
using the extinction law as provided by Cohen et al. (1981) for the $BVRIHKL$
filters and by Koornneef (1983) for the $J$ and $M$ filters.
The visual extinction $A_{\rm V}$ was derived either from 
Neckel \& Klare (1980) with the distance derived from the HIPPARCOS parallax, or
from Burstein \& Heiles (1982) for stars with galactic latitudes $|b| >
10^{\circ}$. In the latter case, the $E_{\rm B-V}$ value provided by Burstein \&
Heiles (1982) was reduced by the factor  
\mbox{$[1 - {\rm exp}(-10 d \sin|b|)]$}, where $d$ stands for the distance in
kpc.
In the remaining cases, the cosecant formula (Feast et al. 1990)
\mbox{$E_{\rm B-V} = 0.032\; ({\rm cosec} |b| - 1)\; [1 - {\rm exp}(-10\; d
\sin|b|)]$} was used.
The adopted $E_{\rm B-V}$ values are listed in Table~\ref{Tab2:sample}.

The deredenned $BVRIJKLM$ magnitudes have then been converted into fluxes using
the zero-magnitude fluxes provided by Johnson (1966). This particular choice will
be justified below.
The zero-magnitude flux in the $H$ band was taken from Jaschek (1978)
and, for the remaining bands, from the original papers quoted above.

A limited sample of 17 stars have enough broad-band colours available ($BVRIJHK$,
as well as IRAS 12, 25 and 60 $\mu$m) to derive the bolometric magnitude by a
direct integration of the available fluxes over wavelength. More precisely, 
the trapezoidal rule has been used on the curve  
$\lambda F_{\lambda}$ versus $\log \lambda$. The zero point of
the bolometric magnitude has been defined from the requirement that $L =
3.86\;10^{33}$
erg~s$^{-1}$ corresponds to \Mbol\ $=4.75$ for the Sun.

If the available photometric data was too scarce to derive the bolometric
magnitude
from a direct integration, it has been derived instead from the $(BC_{\rm K},
V-K)$
relation of Bessell \& Wood (1984) applicable to oxygen-rich stars. 
For the 17 S stars where both methods are applicable, they yield consistent
results (with a r.m.s. deviation of 0.1 mag), provided that the zero-magnitude
fluxes of Johnson
(1966) be adopted (as was done by Bessell \& Wood 1984).  
If the zero-magnitude fluxes listed by Jaschek (1978)
are used instead, somewhat lower bolometric corrections $BC_{\rm K}$ (i.e.
brighter bolometric magnitudes, by about 0.06 mag) are obtained.

A bolometric magnitude \Mbol\ $=-3.74$, based on the same HIPPARCOS parallax, has
recently been derived by van Leeuwen et al. (1997) for $\chi$ Cyg, and agrees 
well with our value $-3.58$ listed in Table~\ref{Tab2:sample}.

\section{The HR diagram of S stars}
\label{section:HR}

\begin{figure}
   \begin{center}
   \leavevmode
   \centerline{\psfig{file=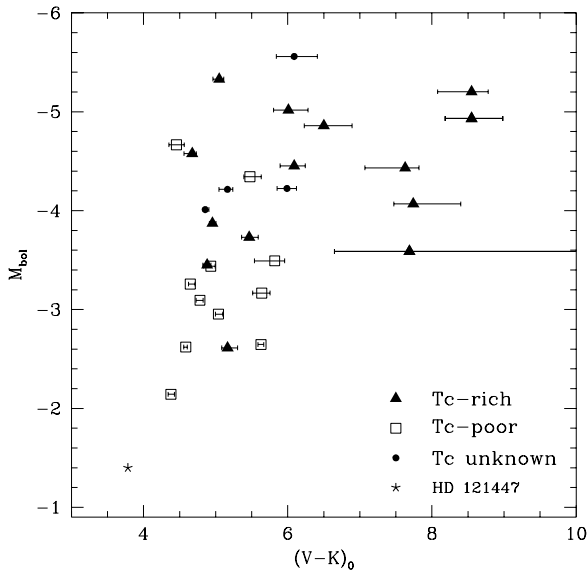,width=8.0cm,height=8.0cm}}
   \end{center}

\caption[]{\label{Fig:HRerrHp}
The HR diagram for S stars with $0 < \sigma_\pi / \pi < 0.85$. Filled triangles
correspond to Tc-rich S stars, open squares to Tc-poor S stars and dots
to S stars with unknown Tc. HD 121447, the boundary case between
Ba and S stars (see Sect.~\ref{Sect:models}), is represented by $\star$.
The error bar provides the uncertainty on $V-K$ caused 
by the intrinsic variability of the HIPPARCOS
\Hp\ magnitude; it covers the range in \Hp\ between the 
5th and 95th percentiles
}
\end{figure}

\begin{figure}
   \begin{center}
   \leavevmode
   \centerline{\psfig{file=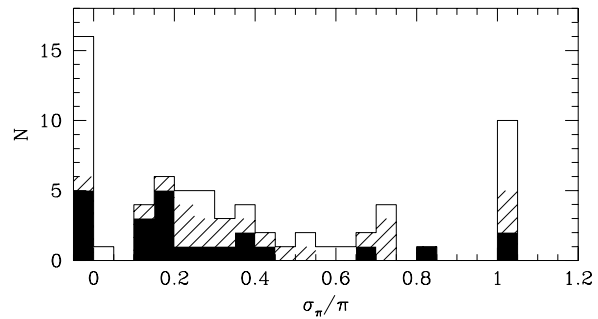,width=8.0cm,rheight=4.0cm}}
   \end{center}
\caption[]{\label{Fig:error}
Distribution of the relative error $\sigma_\pi / \pi$ on HIPPARCOS parallaxes
$\pi$ for S stars.
The hatched and black parts of the histogram correspond to Tc-poor and Tc-rich
S stars, respectively. Stars with negative parallaxes and stars with 
$\sigma_\pi / \pi >0.85$ have been assigned to the leftmost and
rightmost bins, respectively
}
\end{figure}

The $V-K$ colours and the bolometric magnitudes derived as discussed in
Sect.~\ref{Sect:colors}, combined with the HIPPARCOS parallaxes, provide the HR
diagram of S stars presented in Fig.~\ref{Fig:HRerrHp}. Only the 30 S stars with 
an available $K$ magnitude and with $0 < \sigma_\pi / \pi < 0.85$ (see
Fig.~\ref{Fig:error}) have been plotted. 
Despite the sometimes large uncertainties affecting $V-K$ or \Mbol\ 
(as discussed in Sect.~\ref{Sect:errors} below), a segregation
between extrinsic and intrinsic S stars is readily apparent, with {\it extrinsic S stars
being intrinsically fainter and bluer than intrinsic S stars}. 

\subsection{Uncertainties on the HR diagram}
\label{Sect:errors}

The two major sources of uncertainty on the position of an S star in the HR
diagram are the stellar
intrinsic variability and the error on the parallax $\pi$.

The intrinsic variability of S stars has an impact on both $V-K$ and \Mbol.  
The variability in the $V$ band and, to a lesser extent,
in the $K$ band, leads to a variation of $V-K$, and thus of $BC_{\rm K}$
(Sect.~\ref{Sect:BC}).
If $BC_{\rm K}$ was derived from simultaneous $V$ and
$K$ measurements, it could be expected that the variations in $K$ and in
$BC_{\rm K}$  would cancel to a
large extent, thus leaving only a moderate  variation in \Mbol. However, since the $V$
and $K$ data used in this study do not result from simultaneous observations,
the expected cancellation will not occur. Its impact on
\Mbol\ is, however, difficult to evaluate. It seems nevertheless
unlikely that this effect could lead 
to a systematic upwards shift of all intrinsic S stars that would
be responsible for the observed segregation between intrinsic and extrinsic S stars.
The impact of the intrinsic variability 
on $V-K$ is easier to estimate. Its effect is shown on Fig.~\ref{Fig:HRerrHp}
from the variation recorded in the \Hp\ magnitude over the $\sim 1230$~d
duration of the HIPPARCOS mission. The smaller variation due to $K$ has not been
taken into account.  
Clearly, the uncertainty on $V-K$ due to the intrinsic variability does
not jeopardize the observed segregation between extrinsic and intrinsic S stars. 
An interesting by-product of Fig.~\ref{Fig:HRerrHp} is the increase of the amplitude
of variations towards cooler and more luminous stars (see also Eyer
\& Grenon 1997 and Jorissen et al. 1997b).
 
\begin{figure}[t]
   \begin{center}
   \leavevmode
   \centerline{\psfig{file=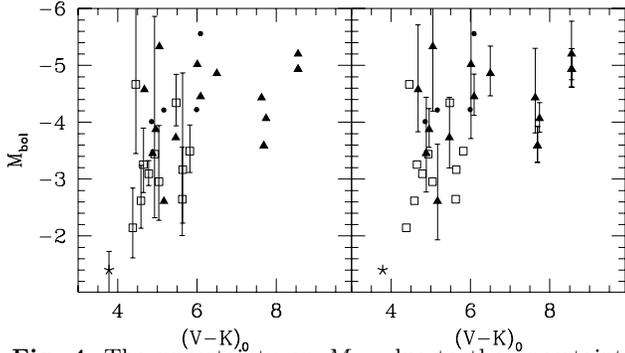,width=9.0cm,rheight=4.0cm}}
   \end{center}
\caption[]{\label{Fig:HRerrpi}
The uncertainty on \Mbol\ due to the uncertainty  $\sigma_\pi$ on the parallax
$\pi$. The error bar extends from \Mbol($\pi + \sigma_\pi)$ to \Mbol($\pi -
\sigma_\pi$). Left panel: parallactic errors for Tc-poor S stars; right panel: 
parallactic errors for Tc-rich S stars.
Symbols are as in Fig.~\ref{Fig:HRerrHp}
}
\end{figure}

The impact on the bolometric magnitude of the uncertainty on the parallax is shown
in 
Fig.~\ref{Fig:HRerrpi}, presenting (separately for intrinsic and extrinsic S
stars) the range of \Mbol\ corresponding to $\pi \pm \sigma_\pi$ (see
Arenou et al. 1995 and Lindegren 1995 for a discussion of the 
external errors of HIPPARCOS parallaxes and of the accuracy of the zero-point).
As discussed by van Leeuwen et al. (1997), there is a specific error source on the
parallax for nearby Mira variables, as some of these stars were found to have
asymmetrical spatial light distributions. Changes in these asymmetries might
affect the derived parallax. The flag 'V' in field H59 of the HIPPARCOS catalogue
possibly reflects such effects, as it refers to a 'variability-induced mover'. 
This flag is set for 14 stars in our sample, mostly nearby Miras (see
column 'Dupl' in Table~\ref{Tab2:sample}). 
Like van Leeuwen et al. (1997), we assume that any such effect will add only 
additional random scatter to the mean results.

Part of the overlap in luminosity between extrinsic and intrinsic S stars in the
HR diagram may actually be attributed to the large error bars
of the interloping stars (NQ Pup and HD 170970\footnote{The presence of Tc in 
HD 170970 is somewhat uncertain, though, since the central wavelength of the
blend containing the $\lambda 426.2$ Tc line lies at the very boundary 
between Tc-rich and Tc-poor stars (Smith \& Lambert 1988)}
among intrinsic S stars, and HD
35155 among the extrinsic S stars). 
That explanation does not hold true, however, for the high-luminosity, 
Tc-poor S star
57 Peg (\Mbol\ $ = -4.3$), which has a small uncertainty on its
parallax ($\sigma_\pi / \pi = 0.21$). That star is special in many respects,
since it has an A6V companion instead of the WD companion expected for
extrinsic S stars in the framework of the binary paradigm
(Sect.~1). It is further discussed in Appendix~A.

The case of HD 35155 deserves further comments. This extrinsic S star
has the second largest relative error on the parallax. It is a binary system  
with an orbital period of $642 \pm 3$~d in a nearly circular orbit
(Jorissen \& Mayor 1992), yielding an
orbital separation of about 2~AU  assuming typical masses of 1.6 and 0.6~\Msun\
for the S star and its suspected white dwarf companion (see Jorissen et al.
1997a). The corresponding angular separation on the sky ($a$) will thus be about twice the
annual parallax (since $a = A \pi$, where $A$ is the orbital separation 
expressed in AU). There is no indication whatsoever that the
orbital motion of HD 35155 has been detected by HIPPARCOS. 
However, since the orbital period is of the order 
of the duration of the HIPPARCOS mission and the parallax is small
($1.32\pm0.99$ mas), this system represented a difficult challenge
for the reduction consortia.
The orbital motion has in fact been detected (flag 'O' in column 'Dupl' 
of Table~\ref{Tab2:sample}) for another
short-period S star (HIC 17296 = HD 22649; $P=596$~d), but the situation is more
favourable in this case, because of its larger parallax ($\pi=6.27$~mas compared
to $\pi=1.32$~mas for HD 35155).

Finally, one should be aware that the observed distribution of
absolute magnitudes of a sample of
stars may be altered by various statistical biases, 
depending on the selection criteria of the sample (e.g. Brown et
al. 1997; Luri \&  Arenou 1997).
Monte-Carlo simulations of Appendix~B
show that the observed segregation between extrinsic and intrinsic S
stars cannot plausibly result from statistical biases altering the true
absolute-magnitude distributions.

\subsection{Comparison with theoretical RGB and AGB evolutionary tracks}
\label{Sect:models}

\begin{figure}[t]
   \begin{center}
   \leavevmode
   \centerline{\psfig{file=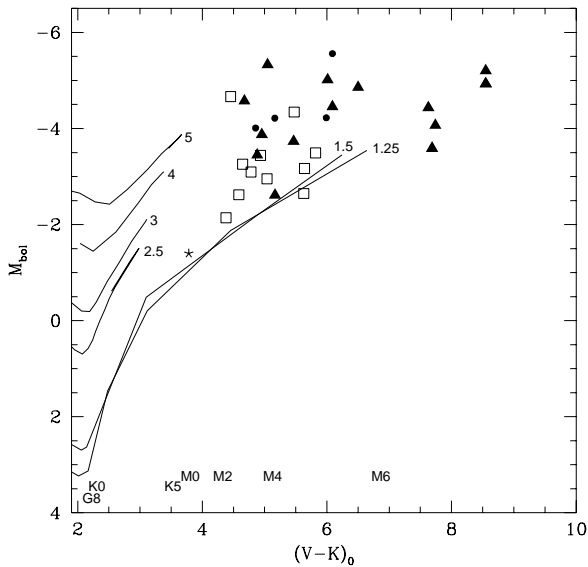,width=8.0cm,rheight=8cm}}
   \end{center}

\caption[]{\label{Fig:HRgenRGB}
The location of the RGB (up to the onset of core He-burning) for stars of various
masses (as labelled, in \Msun) and metallicity $Y = 0.3, Z = 0.02$, according to
Schaller et al. (1992). 
The ($V-K$, spectral types) calibration is from Ridgway et al. (1980).
Other symbols are as in Fig.~\ref{Fig:HRerrHp}
}
\end{figure}

\begin{figure}[t]
   \begin{center}
   \leavevmode
   \centerline{\psfig{file=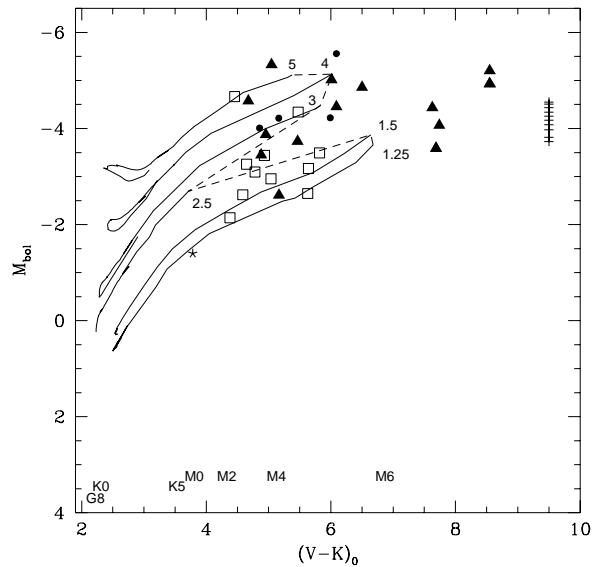,width=8.0cm,rheight=8cm}}
   \end{center}

\caption[]{\label{Fig:HRgenEAGB}
Same as Fig.~\ref{Fig:HRgenRGB}, but for the early AGB up to the first thermal
pulse (Charbonnel et al. 1996). 
To guide the eye, a dashed line connects the starting point of the TP-AGB 
on the various tracks.
The crosses along the right-hand axis provide the luminosities of the first ten
thermal pulses in a 1.5 \Msun, $Z=0.008$ star (corresponding to the LMC
metallicity)  computed by
Wagenhuber \& Tuchman (1996). 
Other symbols are as in Fig.~\ref{Fig:HRerrHp} 
}
\end{figure}

Figure~\ref{Fig:HRgenRGB} compares the position of S stars in the HR diagram with
the RGB (i.e. up
to the onset of core He-burning) for stars of different masses and of metallicity
$Y = 0.3, Z = 0.02$ (Schaller et al. 1992). Figure~\ref{Fig:HRgenEAGB}
is the same as Fig.~\ref{Fig:HRgenRGB}, but for the E-AGB, up to the
first thermal pulse (Charbonnel et al. 1996). 

The effective temperatures of all these models 
have been converted to $V-K$ colours
using the calibration of Ridgway et al. (1980) for class III
giants, which is strictly valid only in the range $2.2<V-K<6.8$.

The predicted location of the giant branch in the HR diagram  
is known to depend sensitively upon model parameters like the convective mixing
length or the atmospheric opacities.
For the Geneva evolutionary tracks used here, these model parameters have been
calibrated so as to reproduce the observed location of the red giant branches
of more than 75 clusters (Schaller et al. 1992). The comparison of these tracks
with the observed location of the S stars in the HR diagram is thus meaningful.
The main result of the present study is apparent on Fig.~\ref{Fig:HRgenEAGB}: the
line marking the onset of thermal pulses matches well the limit between intrinsic
and extrinsic S stars, so that {\it Tc-rich intrinsic S stars may be associated
with thermally-pulsing AGB stars} (the only possible exception being NQ Pup, but
see the discussion about errors below). 

The previous result provides interesting constraints on the occurrence of both the
s-process and the third dredge-up in thermally-pulsing AGB stars, by suggesting
that those processes operate from the very first thermal pulses on, a conclusion
already reached by several authors (e.g., Richer 1981; Scalo
\& Miller 1981; Miller \& Scalo 1982) from the luminosity distribution of carbon
stars in the Magellanic Clouds (see, however, the discussion of
Sect.~\ref{Sect:MC}). Because there is little change in luminosity from one
pulse to the next (see Fig.~\ref{Fig:HRgenEAGB}), and because of the uncertainties
affecting the location of individual S stars in the HR diagram, it is 
difficult, however, to set a limit on the exact 
number of pulses necessary to change
a normal M giant into an (intrinsic) S star.   
Moreover, the statistical biases discussed in Appendix B shift the lower 
boundary of the observed luminosity distribution of Tc-rich S stars 
{\it below} the true threshold.
The Monte-Carlo simulations presented in Appendix~B predict that
the faintest star in the sample of 14 Tc-rich S stars displayed in 
Fig.~\ref{Fig:HRgenEAGB} will be observed 0.7 to 1.5 mag below the
lower boundary of the true luminosity distribution (see Fig.~\ref{Fig:LK}).
This is well in line with the observed location of the Tc-rich S star
NQ Pup (see Sect.~\ref{Sect:errors}) below the TP-AGB threshold luminosity.

As far as extrinsic S stars are concerned, the present data alone 
do not permit to
distinguish between them populating the RGB or the E-AGB of low-mass stars.
Neither does the mass-transfer scenario (Sect.~1) set
any constraint on the {\it current} evolutionary stage of the extrinsic S star.
However, when both the RGB and the E-AGB are possible, evolutionary time-scale
considerations clearly favor the RGB over the AGB.
Besides, the analysis of the orbital elements (Jorissen et al. 1997a) points
towards them being low-mass stars, with an average mass of $1.6\pm0.2$ \Msun. This
value is in excellent agreement with their position in the HR diagram of
Fig.~\ref{Fig:HRgenRGB} (two exceptions are HD 35155 and 57 Peg; see
Sects.~\ref{Sect:errors} and \ref{Sect:57Peg}). 

Contrarily to what might be inferred from the smooth
transition between intrinsic and extrinsic S stars in the HR diagram,
the two kinds of S stars belong to distinct galactic populations,
intrinsic S stars being more concentrated towards the galactic plane
(Jorissen et al. 1993; Jorissen \& Van Eck 1997; Van Eck \& Jorissen, in preparation). 
Extrinsic and intrinsic S stars are thus not simply successive stages 
along the evolution of stars in the same mass range.

Note that the lower left boundary
of the region occupied by extrinsic S stars is set   
by the condition that \Teff\ be low enough in order that  ZrO bands may form. Such
a threshold roughly corresponds to the transition between K and M giants (see the
spectral types labelling Fig.~\ref{Fig:HRgenRGB}), so that extrinsic S stars
should merge into the KIII barium stars at lower luminosities and higher \Teff\
along the RGB (e.g., Jorissen et al. 1997a). As an example, the transition object
HD 121447, classified either  as K7IIIBa5 (L\"u 1991) or S0 (Keenan 1950), has
been located in Fig.~\ref{Fig:HRgenRGB} following the methods presented in
Sect.~\ref{Sect:colors}, using the photometry obtained by Hakkila \& McNamara
(1987) and the HIPPARCOS distance provided by Mennessier et al.
(1997).

\subsection{Comparison with S and carbon stars in external systems}
\label{Sect:MC}

\begin{figure}[t]
   \begin{center}
   \leavevmode
   \centerline{\psfig{file=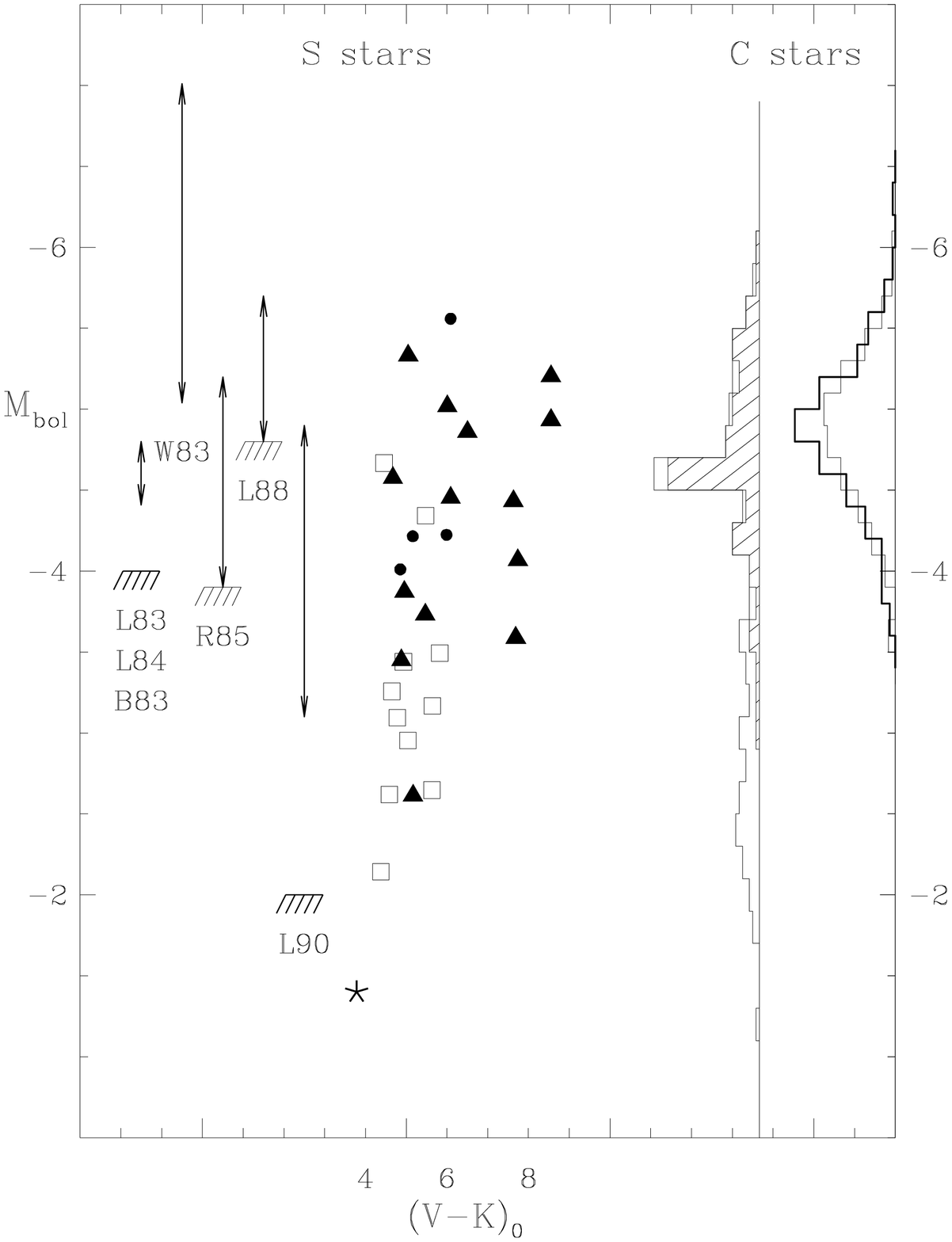,width=8.0cm,height=10.0cm}}
   \end{center}
\caption{
Comparison of the \Mbol\ range of S stars in the
solar neighbourhood (this work, symbols are as in Fig.~\protect\ref{Fig:HRerrHp}) 
and in external systems:\protect\\
\noindent $\bullet$ {\bf S stars}: LMC clusters:
L83 (Lloyd Evans 1983a), L84 (Lloyd Evans 1984), B83 
(Bessell et al. 1983); 
Magellanic Cloud fields:  W83 (Wood et al. 1983), R85 (Reid \& Mould 1985), 
L88 (Lundgren 1988);
Fornax dwarf elliptical galaxy: L90 (Lundgren 1990).
Detection thresholds are also indicated.\protect\\
\noindent $\bullet$ {\bf C stars}: 
the rightmost histogram gives the luminosity function of the 186 LMC (thick line)
and 134 SMC (thin line) C stars identified by Blanco et al. (1980), adopting 18.6 for the
LMC distance modulus and 0.5 mag as the difference in distance 
moduli between the Clouds.
The leftmost histogram provides the luminosity
functions of C stars from the 
Westerlund et al. (1991) SMC survey (hatched) as well as from the deeper  
Westerlund et al. (1995) SMC survey (open)
}
\label{Fig:HRcarbon}
 \end{figure}

Several S stars have been found in the Magellanic Clouds and in a few other nearby
galaxies, allowing a direct comparison of their luminosities with those of 
galactic S stars derived from the HIPPARCOS trigonometric parallaxes
(Fig.~\ref{Fig:HRcarbon}):
\begin{itemize}
\item[$\bullet$]
In a study of 90 long-period variables in the Magellanic Clouds, Wood et al.
(1983) identified 14 MS stars (labelled as W83 on
Fig.~\ref{Fig:HRcarbon}) in the range 
$-7 \le $\Mbol\ $\le -5$;
\item[$\bullet$]
In a magnitude-limited survey of several fields in the outer regions
of the northern LMC, Reid \& Mould (1985) identified 10 S stars (labelled as R85
on Fig.~\ref{Fig:HRcarbon}) in the range 
$-5.0 \le $\Mbol\ $\le -3.9$;
\item[$\bullet$]
In a sample extracted from the faint tail of the Westerlund et
al. (1981) LMC survey of late-type giants,
Lundgren (1988) finds 6 S stars (labelled as L88 on
Fig.~\ref{Fig:HRcarbon}) in the range $-5.7 \le M_{\rm bol} \le -4.8$;
\item[$\bullet$]
Bessell et al. (1983) and Lloyd Evans (1983a, 1984) find  
16 S stars  (labelled as B83, L83 and L84 on Fig.~\ref{Fig:HRcarbon},
with bolometric magnitudes provided in  
Tables A1 and A2 of Westerlund et al. 1991) in 
LMC clusters with $-4.8 \le M_{\rm bol} \le -4.4$
(excluding NGC 1651/3304, quite far from the cluster center, possibly
a field star, and the close pair LE1+2 in NGC 1987);
\item[$\bullet$]
S stars have also been found in other galaxies: for example, 
Aaronson et al. (1985) find one S star 
with $M_{\rm bol}=-5.15$ in NGC 6822, Brewer et al. (1996)
discover one S star with $M_{\rm bol}=-5.3$ in M31, and
Lundgren (1990) finds seven S stars in the range $-4.9 \le M_{\rm bol} \le -3.1$ 
in the Fornax dwarf elliptical galaxy (labelled as L90 on Fig.~\ref{Fig:HRcarbon}).
\end{itemize}

The bolometric-magnitude range for all these S stars in external systems has been plotted
in Fig.~\ref{Fig:HRcarbon}. Their luminosities are generally comparable to those
of the galactic Tc-rich S stars, with the exception of the brighter W83 S stars.
These W83 S stars extend up to the theoretical AGB tip corresponding to the 
Chandrasekhar limit for the degenerate core. Smith et al. (1995) have shown that,
in the Magellanic Clouds, these S stars with $-7 \le $ \Mbol $\le -6$ are all Li-
strong stars with $M \ga 4$~\Msun.  
In the solar neighbourhood, these Li-strong S stars are rare (Catchpole \& Feast
1976; Lloyd Evans \& Catchpole 1989). 
The only such star in the present HIPPARCOS sample is T Sgr, but its
HIPPARCOS parallax is useless (Table~\ref{Tab1:sample}).

Conversely, in all the external systems, there appears to be a lack of
low-luminosity S stars with respect to the solar neighbourhood. 
In the Magellanic Cloud fields surveyed, this lack of low-luminosity S
stars may clearly be attributed to the limited sensitivity of the surveys 
(see R85 and L88 in Fig.~\ref{Fig:HRcarbon}). 
The situation is different in the Magellanic Cloud clusters, where 
the available surveys are sensitive down to \Mbol\ $= -4.0$ and find
many M stars but no S stars in the range $-4.0$ to $-4.4$.    
Does this mean that Tc-poor, extrinsic S stars are really absent from
the Magellanic Cloud globular clusters? Possibly, although extrinsic
S stars are mostly found at luminosities fainter than \Mbol\ $=-3.5$
(i.e. below the RGB-tip), and were thus not properly surveyed
in the Magellanic Cloud clusters. The absence of extrinsic S stars
in the Magellanic Clouds clusters would not be surprising, though,
given the situation encountered in galactic globular clusters.
Barium and (low-luminosity) S stars are rare in galactic
globular clusters (Vanture et al. 1994, and
references therein). They are only present in the massive,
low-concentration cluster $\omega$ Cen (Lloyd Evans 1983b),
and their origin in that cluster is still unclear. 
C\^ot\'e et al. (1997) have argued that the binary
evolution leading to the formation of extrinsic heavy-element-rich
stars is only possible in low-concentration clusters like $\omega$
Cen. In more concentrated clusters, hard binaries
rapidly shrink to orbital separations not large enough to accommodate an AGB
star. It would be of interest to check whether Magellanic Cloud
globular clusters are indeed too concentrated to allow the formation
of extrinsic S stars. However, the recent result that the barium and
the low-luminosity S stars in $\omega$ Cen appear to have constant 
radial velocities (Mayor et al. 1996) challenges the above picture. It
may indicate that the barium and low-luminosity S stars in
$\omega$ Cen represent the extreme tail of the wide range in metal
abundances observed in that particular globular cluster, as suggested
by Lloyd Evans (1983b). Alternatively, these stars might have been
enriched in the past by mass transfer in soft binary systems which
were later dynamically disrupted.
 
The lack of low-luminosity S stars in the Fornax dwarf elliptical
galaxy (L90 in Fig.~\ref{Fig:HRcarbon}) is probably related to the
low-metallicity of that system ([Fe/H] $\sim -1.5$; Lundgren
1990). At such low metallicities, the red giant branch is shifted
towards the blue, so that stars with luminosities typical of the
galactic Tc-poor S stars are actually of spectral type G or K in Fornax
(see Smith et al. 1996, 1997 for a discussion on a similar situation in the
galactic halo). No specific effort to find heavy-element-rich stars
among the warm (`continuum') 
giants  was attempted by Lundgren (1990). Contrarily to the situation
prevailing in the Magellanic Cloud clusters discussed above, an S star
is found in Fornax at the luminosity threshold (\Mbol\ $= -3.2$)   
between `continuum' (G or K) giants and M giants, suggesting that the
lower luminosity cutoff observed for S stars in Fornax is just 
a spectral selection effect.    

Deeper surveys are available for carbon stars, because of their more
conspicuous spectral features. 
The luminosity distribution of the 134 SMC and 186 LMC carbon stars identified by
the pioneering survey of Blanco et al. (1980) is shown in
Fig.~\ref{Fig:HRcarbon}. It should be noted that the luminosity functions are
almost identical for the LMC and SMC despite their different metallicities, so
that the comparison with a more metal-rich galactic sample is not unreasonable.
A GRISM survey of carbon stars in the SMC by Rebeirot et al. (1993; RAW) led to
the discovery  of 1707 such stars.
For 100 of them, bolometric magnitudes were determined from $JHK$ photometry
(Westerlund et al. 1991), with a
limiting magnitude of $M_{\rm bol} \sim -3$. Their distribution is represented by
the hatched histogram on Fig.~\ref{Fig:HRcarbon}. 
A subsequent photometric and spectroscopic survey has been devoted 
to the $\sim 5\%$ among 
RAW objects having $M_{\rm bol}>-3$  (Westerlund et al. 1995;
Fig.~\ref{Fig:HRcarbon}). 
{\it The range of absolute bolometric magnitudes for carbon
stars from the deeper SMC survey now totally covers that of galactic S stars,
including its low-luminosity tail.}

The nature of the SMC low-luminosity carbon  stars is still debated.
They might either be contaminating {\it dwarf} carbon (dC) stars from our own
Galaxy,
or low-luminosity carbon stars equivalent to the galactic R carbon stars, or may
be
extrinsic carbon stars formed by mass transfer across a binary system
like galactic extrinsic S stars
(as already suggested by Barnbaum \& Morris 1993).
Westerlund et al. (1995) reject the first hypothesis, 
mainly because none (except two)
of their carbon stars have colours similar to the known galactic dC stars, and
moreover, none
exhibits a strong C$_2$ band head at 619.1 nm, a feature exhibited
by galactic dC stars.
The last two hypotheses offer interesting alternatives that remain to be
investigated. It should also be noted  that the possibility that some of the low-
luminosity SMC carbon stars be extrinsic carbon stars 
makes it difficult to use the SMC carbon-star luminosity distribution to derive 
the luminosity threshold for 
the occurrence of the s-process and third dredge-up along the AGB (see the
discussion in Sect.~\ref{Sect:models}). Further observations are clearly required
in order to distinguish intrinsic carbon stars from possible extrinsic carbon
stars.

\section{Infrared excess and position along the giant branches}
\label{Sect:IR}

Infrared excesses revealing the presence of
cool circumstellar material are a common feature
of late-type giants (e.g., van der Veen \& Habing 1988; Habing 1996).
These excesses are often associated with intrinsic stellar variability. Hacking 
et al. (1985) have shown that the brightest IRAS $12\mu$m sources outside
the galactic plane are long-period variables (LPV). Similarly, in globular
clusters, all stars with a 10 $\mu$m excess are LPVs (Frogel \& Elias 1988), and
moreover, they are found only above the RGB tip.
These IR excesses are associated with a strong mass loss, which thus appears to
be the rule among LPVs. It is expected to be much smaller in non-pulsating AGB and
RGB stars (e.g., Habing 1987).
Consequently, S stars 
with large photometric variability and infrared excesses are expected
to be found among the most luminous stars.
Indeed, it was already shown in Fig.~\ref{Fig:HRerrHp} that variability in the
HIPPARCOS \Hp\ band tends to increase with luminosity. 
The present data offer the possibility to investigate as well the evolution of
the IR excesses probing mass loss along the giant branches in the HR diagram.  

\begin{figure}[t]
   \begin{center}
   \leavevmode
   \centerline{\psfig{file=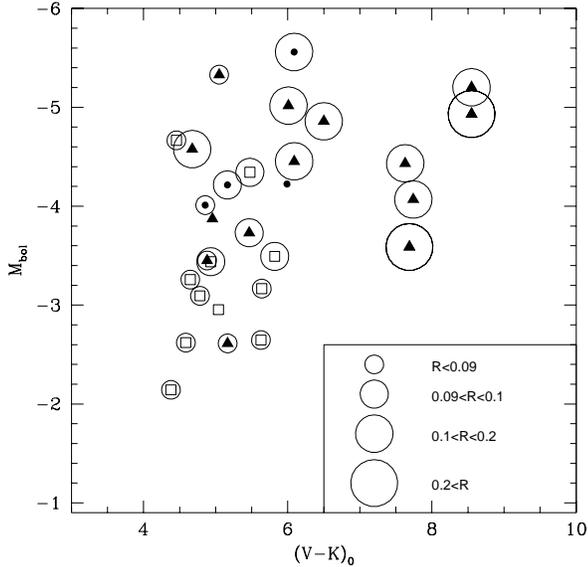,width=8.0cm,height=8.0cm}}
   \end{center}
\caption[]{\label{Fig:HRiras}
Infrared excess along the giant branches,
as measured by the ratio $R=F(12\mu{\rm m})/F(2.2\mu{\rm m})$. The dichotomy
between extrinsic and intrinsic S stars is clearly apparent.
Symbols are as in Fig.~\ref{Fig:HRerrHp}
}
\end{figure}

The ratio $R=F(12\mu{\rm m})/F(2.2\mu{\rm m})$, probing the presence of cool
circumstellar material emitting at 12 $\mu$m, has been derived from the IRAS PSC
and from the $K$ magnitude, whenever available
(using the calibration of Beckwith et al. 1976 to convert $K$ magnitudes into
fluxes at 2.2 $\mu$m). As already shown by Jorissen et al. (1993), 
there is a clear segregation of S stars according to their value of $R$. All
Tc-deficient S stars have $0.075 \la R \la 0.093$,
consistent with photospheric blackbody colours
($R=0.073$ for \Teff = 4000~K and $R = 0.093$ for \Teff= 3150~K, which corresponds
to the \Teff\ range spanned by extrinsic S stars).
On the contrary, intrinsic S stars generally have $R > 0.1$, indicative of
circumstellar
dust. This segregation is a further indication of the inhomogeneous
nature of the family of S stars.

As expected, Fig.~\ref{Fig:HRiras} clearly shows that $R < 0.1$ indices are
restricted to the RGB (or E-AGB), whereas S stars on the AGB have larger and larger $R$
indices as they evolve towards cooler \Teff.

\section{Conclusions}

All stars with ZrO bands were historically assigned to a unique spectral
class (S). 
However, an increasing number of arguments recently suggested that the
S family comprises stars of two different kinds: Tc-poor binary (extrinsic) S
stars and Tc-rich (intrinsic) S stars.
The HR diagram of S stars constructed from HIPPARCOS
parallaxes presented in this study 
fully confirms that dichotomy, by showing that: 
\begin{itemize}
\item [$\bullet$] extrinsic S stars are hotter and intrinsically fainter 
than intrinsic S stars;

\item [$\bullet$] extrinsic S stars are low-mass stars populating either the RGB
or the E-AGB, and in any case are found {\it below} the luminosity threshold
marking the onset of thermal pulses on the AGB. Therefore,  their
chemical peculiarities cannot originate from 
internal nucleosynthesis and dredge-up processes occurring along the TP-AGB. Their
binary nature suggests instead that their chemical peculiarities originate from
mass transfer;

\item [$\bullet$] on the contrary, Tc-rich S stars are found just above the TP-AGB
luminosity threshold. Their location in the HR diagram indicates that s-process
nucleosynthesis and third dredge-up must be operative quite early on the TP-AGB.

\end{itemize}

The galactic latitude distributions of extrinsic and intrinsic S stars are also clearly
different (Jorissen et al. 1993; Jorissen \& Van Eck 1997), 
indicating that they are not simply successive stages along the evolution of stars
in the same mass range.

\acknowledgements{We would like to thank F. Pont for useful discussions. 
S.V.E. is supported by a F.R.I.A. (Belgium) doctoral fellowship, and 
A.J. is Research Associate (F.N.R.S., Belgium). Financial support from the {\it Fonds
National de la Recherche Scientifique} (Belgium, Switzerland) is gratefully acknowledged.  
This research has made use of the {\it General Catalogue of Photometric Data},
operated at the Institute of Astronomy, University of Lausanne (Switzerland) and
of the SIMBAD database, operated at CDS, Strasbourg (France).
}

\renewcommand{\thesection}{Appendix \Alph{section}}
\setcounter{section}{0}
\renewcommand{\thefigure}{\Alph{section}\arabic{figure}}
\setcounter{figure}{0}

\section{57 Peg: A very luminous extrinsic S star}
\label{Sect:57Peg}

\begin{figure}[t]
   \begin{center}
   \leavevmode
   \centerline{\psfig{file=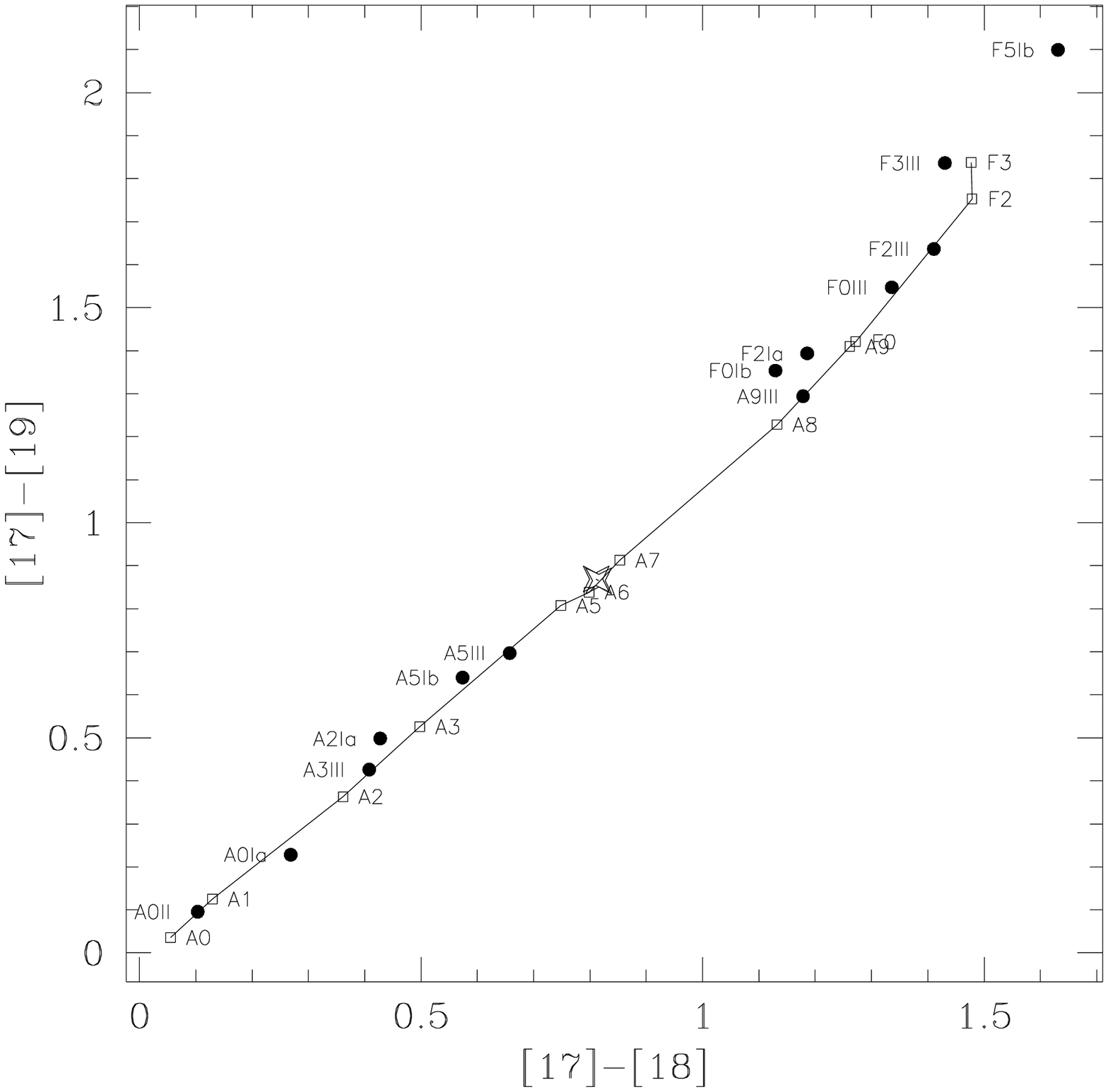,width=8.0cm,rheight=8cm}}
   \end{center}

\caption[]{\label{Fig:IUE}
The $([17]-[18], [17]-[19])$ colour-colour diagram for 57 Peg (cross) and standard
stars from the spectral atlas of Wu et al. (1983). Main sequence stars are
represented by open squares with labels on the right giving the corresponding {\it
optical} spectral type. Giants and supergiants are represented by filled circles,
with labels on the left of the corresponding point  
}
\end{figure}

The star 57 Peg is peculiar in being a high-luminosity extrinsic S star (\Mbol $
= -4.3$; Table~\ref{Tab2:sample}). Its outlying position cannot be attributed 
to the uncertainty on its parallax, which is small ($\sigma_\pi/\pi = 0.21$;
Fig.~\ref{Fig:HRerrpi}).
It is a long-period binary system ($P > 3700$~d; Jorissen et al. 1997a) with a
composite spectrum caused by a companion of spectral type approximately A3V, as
derived by Hackos \& Peery (1968) from the analysis of a violet optical spectrum.
The A-type main sequence companion is much better visible on the IUE spectra
available in the archive (Peery 1986). 
In order to derive its exact spectral class, and thus its mass, its UV colours were
compared to standard stars  from the atlas of Wu et al. (1983).
Figure~\ref{Fig:IUE} presents the location of 57 Peg in the $([17]-[18], [17]-
[19])$ colour-colour diagram, where $[i] - [j] = -2.5 \log (F_i/F_j)$ and $F_i$ is
the average flux in the spectral ranges 165.5 -- 175.5 nm (for $i = 17$), 175.5
-- 185.5 nm ($i = 18$) or 185.5 -- 195.5 nm ($i = 19$). All the stars from the
atlas of Wu et al. (1983) plotted on Fig.~\ref{Fig:IUE} have a flux well
above the noise level in all three bands. For 57 Peg, the two spectra used
(SWP05384, recorded on May 28, 1979 and SWP32554, obtained on December 18, 1987) 
yield  identical colours, despite the fact that they were obtained about 8 years
apart. 
All spectra were retrieved from the final IUE archive processed with the NEWSIPS
software. No reddening correction has been applied, since it is assumed to be
negligible given the small colour excesses listed by Wu et al. (1983) and by Hackos
\& Peery (1968).

Main sequence stars fall along a straight line in Fig.~\ref{Fig:IUE}, with 57 Peg
falling around spectral type A6. Although the colour-colour diagram does not probe
efficiently the luminosity class, the individual fluxes of 57 Peg are compatible
with the companion being an A6 star on the main sequence, since 
its fluxes match those of the A6V standard HD 28527 (SWP19459), assuming $M_{\rm v} =
+2.05$ for A6V stars, and given the distance derived for the 57 Peg system from
the HIPPARCOS parallax. The corresponding mass for the companion is then 1.9
\Msun. The S star primary must have evolved faster and should thus be more massive
than 1.9 \Msun, which is consistent with its position in the HR diagram along the
3~\Msun\
E-AGB (Fig.~\ref{Fig:HRgenEAGB}). That S star is thus likely to be more massive
than the average extrinsic S star (for which $\langle M \rangle=1.6\pm0.2$ \Msun; 
Sect.~\ref{Sect:models} and Jorissen et al. 1997a). 

The Tc-poor nature of 57 Peg is quite puzzling,
though, since the binary paradigm (Sect.~1) requires the companion of 
a Tc-poor S star to be a WD rather than a main sequence star. 
Possibilities to resolve this puzzle within the
framework of the binary paradigm for extrinsic S stars include (i) 
57 Peg is a triple system (S+A6V+WD), or (ii) the companion is an
accreting WD mimicking a main sequence spectrum, or (iii) 57 Peg is not an S
star at all. The radial-velocity data accumulated
so far does not allow us to test possibility (i) yet. 
Possibility (ii) is unlikely, since the SWP spectrum carries
no sign of binary interaction (like the CIV $\lambda 155.0$ doublet), the
two spectra taken 8 years apart yield identical fluxes (whereas some
variability is expected in the case that the observed continuum be
due to accretion on a WD), and  
finally, in the regime of rapid mass accretion by a WD, where a stable
H-burning shell forms, the accreting WD mimics a supergiant
rather than a main sequence star (e.g., Paczy\'nski \& Rudak 1980). 
Possibility (iii) - 57 Peg is not an S star - has already been
mentioned explicitly by Smith \&
Lambert (1988) with reference
to an earlier paper (Smith \& Lambert 1986), although 57 Peg is not to be found
in that other paper. This last possibility clearly deserves a closer study.  
Otherwise, 57 Peg would add to the small set of Tc-poor S stars (HD 191589, HDE
332077) with a main sequence companion (Jorissen \& Mayor 1992; Ake \& Johnson
1992; Ake et al. 1994; Jorissen et al. 1997a). 

\section{An evaluation of the impact of statistical biases}
\renewcommand{\theequation}{B.\arabic{equation}}
\setcounter{figure}{0}

The evaluation of the statistical biases altering
the true absolute magnitude distribution of any given
observed sample of stars requires the knowledge
of the selection criteria defining that sample. 
Indeed, magnitude-limited samples are mostly affected by the Malmquist bias
(Malmquist 1936), whereas the Lutz-Kelker bias plays an important role for
parallax-limited samples (e.g. Lutz \& Kelker 1973; Lutz 1979; 
Hanson 1979; Smith 1988).
For the HIPPARCOS sample of S stars considered here, 
the selection criteria combine limits on the parallax 
with limits on the magnitude. 
Monte-Carlo simulations (see e.g. Pont et al. 1997) appear
therefore more appropriate than an analytical
approach to explore the biases resulting from the selection criteria and
the observation errors when applied to a reasonable
model of the parent population.

The main aim of the simulations presented in this Appendix 
is to demonstrate that the segregation observed in the HR
diagram between Tc-rich and Tc-poor S stars is not an artefact
caused by statistical biases which might in principle operate
differentially on the two families (because e.g. of different
galactic distributions), but that it must result instead 
from truly different luminosity distributions.

To this aim, a parent population of {\it Tc-rich} S stars is generated 
with the following properties:

\begin{itemize}
\item[$\bullet$] a large number ($>300~000$) of Tc-rich S stars are created,
with a spatial distribution characterized by
an exponential scale height of 180~pc 
(as derived for the Tc-rich stars of the Henize sample 
comprising 205 S stars; Van Eck \& Jorissen, in preparation), and
a uniform projected distribution on the galactic 
plane around the Sun;
\item[$\bullet$] various bolometric-magnitude distributions
are adopted (see below);
\item[$\bullet$] the $(V-I)_{\rm J}$ colour index 
is derived from the bolometric magnitude using a fiducial AGB
fitted to our sample:
\begin{equation}
(V-I)_{\rm J} = (0.5 - $ \Mbol\ $ )/1.5 + \sigma_{\rm {V-I}}~$gauss$(0,1)
\label{Eq:AGB}
\end{equation}
where $\sigma_{\rm {V-I}} = 0.6$ and gauss(0,1) is a random variable 
with a reduced normal distribution;
\item[$\bullet$] the bolometric correction is computed from the
$(BC_{\rm I},(V-I)_{\rm J})$ relation of Bessell \& Wood (1984);
\item[$\bullet$] the (unreddened) $V_{\rm J,0}$ magnitude is deduced from 
the distance $d$, \Mbol, $BC_{\rm I}$ and $(V-I)_{\rm J}$;
\item[$\bullet$] the $V_{\rm J,0}$ magnitude is reddened according
to the cosecant formula (Feast et al. 1990).
\end{itemize}

Stars are extracted from this parent population so as to reproduce
the {\it observed} distribution of $V_{\rm J}$ magnitudes in the
entire\footnote{
The $V_{\rm J}$ distribution of the entire HIPPARCOS
sample of S stars (60 stars when excluding the two close 
visual binaries and the misclassified S star; see 
Table~\ref{Tab1:sample}) has been preferred over the $V_{\rm J}$
distribution of the 21 S stars known to be Tc-rich, because
of its larger statistical significance and because the observed
$V_{\rm J}$ distributions of Tc-rich and Tc-poor S stars are not
significantly different.} HIPPARCOS sample of S stars.
The ``measured'' parallax of each extracted star is then computed
from its distance $d$ by 
\begin{equation}
\pi_{\rm measured} = 1/d + \sigma_{\pi}(V_{\rm J})~$gauss$(0,1)
\end{equation}
where the $(\sigma_{\pi},V_{\rm J})$ relationship is derived 
from a least square fit to the HIPPARCOS data for S stars.
The simulated star is then retained if it satisfies the 
condition  $ 0 < \sigma_{\pi}/\pi < 0.85$ imposed on the star
in order to be included in the HR diagram of Fig.~\ref{Fig:HRerrHp}.

\medskip
In summary, the two selection criteria are:
\begin{itemize}
\item Condition I: the $V_{\rm J}$ distribution of the 
extracted sample of simulated stars
has to reproduce the $V_{\rm J}$ distribution of the HIPPARCOS
sample of S stars;
\item Condition II: $ 0 < \sigma_{\pi}/\pi < 0.85$.
\end{itemize}

The stars extracted from the parent population and satisfying 
condition I (resp. I+II) define what will be called in the
following the sample $\Sigma_{\rm I}$ (resp. $\Sigma_{\rm I+II}$).
The sample $\Sigma_{\rm I}$ is supposed to represent the real
HIPPARCOS sample.

Although the selection criteria that were used to include
a given S star in the HIPPARCOS Input Catalogue are unknown
to us, condition I ensures
that they are implicitly met in our simulation.
Besides, condition II is satisfied by 61\% of the stars 
in $\Sigma_{\rm I}$ to be compared
with 63\% in the real HIPPARCOS sample of S stars.
This good agreement constitutes a check of the internal consistency
of our model as a whole.

\begin{figure}
   \begin{center}
   \leavevmode
   \centerline{\psfig{file=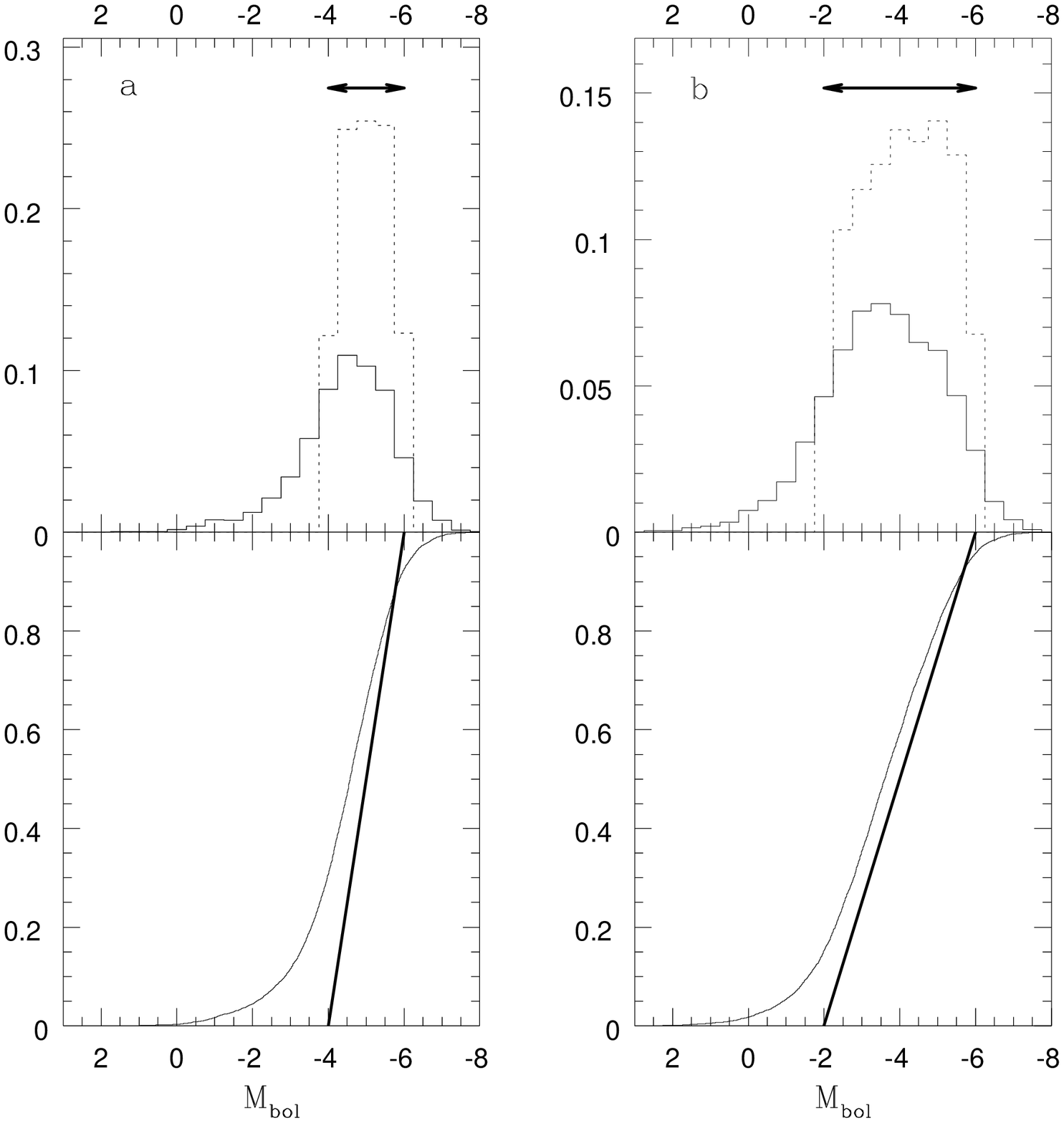,width=8.0cm,height=8.0cm}}
   \end{center}

\caption[]{\label{Fig:LK}
Bolometric magnitude distributions (upper panels) and their
cumulative frequencies (lower panels) resulting from the Monte-Carlo
simulations described in the text.
The {\it parent} \Mbol\ distributions are rectangular and span the
range indicated by the arrows in the upper panels.
In the lower panel, their cumulative frequency is represented by
the thick solid line.
The dotted line (resp. thin solid line) refers to the distribution
of the $\Sigma_{\rm I}$ (resp. $\Sigma_{\rm I+II}$) sample.
Both histograms are drawn at the same scale, the dotted histogram
being normalized to unity.
Note that a Malmquist-type bias is responsible for the deviation of the
dotted histogram from the original rectangular distribution,
whereas a Lutz-Kelker-type bias causes the difference
between the dotted- and the thin-solid histograms
}
\end{figure}

The comparison between the \Mbol\ distributions of the 
sample $\Sigma_{\rm I+II}$ and of the parent sample then
illustrates the impact of the statistical biases. This is shown
in Fig.~\ref{Fig:LK} for rectangular parent distributions in the range
$-4 \ge $ \Mbol\ $ \ge -6$ (Fig.~\ref{Fig:LK}a) and in the range
$-2 \ge $ \Mbol\ $ \ge -6$ (Fig.~\ref{Fig:LK}b).
The net effect of these biases is to transform the original
rectangular distribution into a bell-shaped distribution,
with a median  shifted by 0.4 magnitude towards fainter
objets, and with asymmetrical tails extending beyond the original
limits. This spread is most pronounced on the fainter side of the
distribution. 
The faintest {\it observed} S star in a sample of $N$ objects
must therefore be expected beyond the lower boundary of the
{\it parent} distribution (at a bolometric magnitude corresponding
to the cumulative frequency $1/N$). The amplitude of this offset does
not depend on the median value of the parent distribution,
but depends rather sensitively upon its width (compare 
Fig.~\ref{Fig:LK}a and Fig.~\ref{Fig:LK}b). For reasonable values of the
width of this parent \Mbol\ distribution, and adopting
$N=14$ as for the sample of Tc-rich S stars plotted in
the HR diagram of Fig.~\ref{Fig:HRerrpi}, 
the offset value is found to lie between 0.7 and 1.5 mag.
According to this statistics, 
the presence of the Tc-rich S star NQ Pup at \Mbol\ $=-2.6$
(i.e. about 1 mag below the onset of the TPAGB) is
nevertheless fully compatible with its location on the TPAGB 
(see Fig.~\ref{Fig:HRgenEAGB}).
Furthermore, Fig.~\ref{Fig:LK}b demonstrates {\it ad absurdum}
that the luminosity segregation between Tc-rich and Tc-poor 
S stars is not an artefact: if Tc-rich S stars with
$-2 \ge $ \Mbol\ $ \ge -4$ typical of Tc-poor S stars
were to exist, they should have been detected indeed.

Two conclusions may thus be drawn from the above simulations:
\begin{itemize}
\item they demonstrate that the lack of
low-luminosity Tc-rich S stars
does not result from statistical biases, since these biases
tend to {\it increase} the number of low-luminosity stars;

\item when inferring the luminosity threshold for the appearance
of Tc-rich S stars, it must be reminded that statistical biases
tend to make the faintest Tc-rich S stars observed by HIPPARCOS 
appear {\it below} (between 0.7 and 1.5 mag)
the true lower-luminosity boundary of the parent population.
\end{itemize}


\begin{thebibliography}{}

\bibitem[]{}
Aaronson M., Mould J., Cook K.H., 1985, ApJ 291, L41

\bibitem[]{}
Ake T.B., Johnson H.R., 1992. In:  Giampapa M.S., 
Bookbinder J.A. (eds.) Seventh Cambridge Workshop on Cool
Stars, Stellar Systems and the Sun. Astron. Soc. Pacific Conf. Series
26, p. 579

\bibitem[]{}
Ake T., Jorissen A., Johnson H.R., Mayor M., Bopp B., 1994, 
Bull. Am. Astron. Soc. 24, 1280  

\bibitem[]{}
Arenou F., Lindegren L., Froeschle M., Gomez A.E., Turon C.,
Perryman M.A.C., Wielen R., 1995, A\&A 304, 52

\bibitem[]{}
Barnbaum C., Morris M., 1993, Bull. American Astron. Soc., 182, \#46.17

\bibitem[]{}
Beckwith S., Evans N.J., Becklin E.E., Neugebauer G., 1976, 
ApJ 208, 390

\bibitem[]{}
Bessell M.S., Wood P.R., Lloyd Evans T., 1983, MNRAS 202, 59

\bibitem[]{}
Bessell M.S., Wood P.R., 1984, PASP 96, 247

\bibitem[]{}
Bidelman W.P., Keenan P.C., 1951, ApJ 114, 473

\bibitem[]{}
Blanco V.M., McCarthy M.F., Blanco B.M., 1980, ApJ 242, 938

\bibitem[]{}
Boesgaard A.M., 1969, PASP 81, 283

\bibitem[]{}
B\"{o}hm-Vitense E., Nemec J., Proffitt C., 1984, ApJ 278, 726

\bibitem[]{}
Brewer J.P., Richer H.B., Crabtree D.R., 1996, AJ 112, 491

\bibitem[]{}
Brown J.A., Smith V.V., Lambert D.L., Dutchover E.Jr., Hinkle K.H.,
Johnson H.R., 1990, AJ 99, 1930
 
\bibitem[]{}
Brown A.G.A., Arenou F., van Leeuwen F., Lindegren L., Luri X., 
1997. In: Perryman M.A.C. (ed.) The Hipparcos and
Tycho Catalogues, ESA-SP 402, in press

\bibitem{} 
Burstein D., Heiles C., 1982, AJ87, 1165

\bibitem{} 
Catchpole R.M., Feast M.W., 1976, MNRAS 175, 501

\bibitem{} 
Catchpole R.M., Robertson B.S.C., Lloyd Evans T.H.H., Feast
M.W., Glass I.S., Carter B.S., 1979, SAAO Circ. 1, 61

\bibitem[]{}
Charbonnel C., Meynet G., Maeder A., Schaerer D., 1996, A\&AS 115, 339 

\bibitem[]{}
Chen P. S., Gao H., Chen Y. K., Dong H. W., 1988, A\&AS 72, 239

\bibitem[]{}
Cohen J.G., Frogel J.A., Persson S.E., Elias J.H., 1981, ApJ 249, 481,

\bibitem[]{}
C\^ot\'e P., Hanes D.A., McLaughlin D.E., Bridges T.J., Hesser J.E.,
Harris G.L.H., 1997, ApJ 476, L15
 
\bibitem[]{}
Culver R.B., Ianna P.A., 1975, ApJ 195, L37

\bibitem[]{}
Eggen O.J., 1972a, PASP 84, 406

\bibitem[]{}
Eggen O.J., 1972b, ApJ 177, 489

\bibitem[]{}
ESA, 1997. The HIPPARCOS Catalogue, ESA SP-1200
 
\bibitem[]{}
Eyer L., Grenon M., 1997. In: Perryman M.A.C. (ed.) The Hipparcos and
Tycho Catalogues, ESA-SP 402, in press

\bibitem[]{}
Feast M.W., 1953, MNRAS 113, 510

\bibitem[]{}
Feast M.W., Catchpole R.M., Glass I.S., 1976, MNRAS 174, 81P

\bibitem[]{}
Feast M.W., Robertson B.S.C., Catchpole R.M., Lloyd Evans T.,
Glass I.S., Carter B.S., 1982, MNRAS 201, 439

\bibitem[]{}
Feast M.W., Whitelock P.A., Carter B.S., 1990, MNRAS 247, 227

\bibitem[]{}
Fouqu\'e P., Le Bertre T., Epchtein N., Guglielmo F., 
Kerschbaum F., 1992, A\&AS 93, 151

\bibitem[]{}
Frogel J.A., Elias J.H., 1988, ApJ 324, 823

\bibitem[]{}
Gillett F.C., Merrill K.M., 1971, ApJ 164, 83

\bibitem[]{} 
Habing H.J., 1987. In:
Appenzeller I.,  Jordan C. (eds.) Circumstellar Matter (IAU Symp. 122).  Reidel,
Dordrecht, p.197

\bibitem[]{}
Habing H.J., 1996, ARA\&A 7, 97

\bibitem[]{}
Hacking P., Neugebauer G., Emerson J., et al., 1985, PASP 97, 616

\bibitem[]{}
Hackos W.Jr., Peery B.F.Jr., 1968, AJ 73, 504

\bibitem[]{}
Hakkila J., McNamara B.J., 1987, A\&A 186, 255

\bibitem[]{}
Hanson R.B., 1979, MNRAS 186, 875

\bibitem[]{}
Henize K.G., 1960, AJ 65, 491

\bibitem[]{}
Houk N., Cowley A.P., 1975. The Michigan Catalogue of Two-Dimensional Spectral
Types for the HD Stars. Univ. Michigan, Ann Arbor, Vol. 1

\bibitem[]{}
Iben I.Jr., Renzini A., 1983, ARA\&A 21, 271

\bibitem[]{}
IRAS Science Team, 1988. IRAS Point Source Catalogue, prepared
by Beichman C.A., Neugebauer G., Habing H.J., Clegg P.E.,
Chester T.J., NASA-RP 1190

\bibitem[]{}
Jaschek C., 1978, Bull. Inf. CDS 15

\bibitem[]{}
Jaschek C., Jaschek M., 1987. The Classification of Stars, Cambridge Univ. Press

\bibitem[]{}
Johnson H.L., 1966, ARA\&A 4, 193

\bibitem[]{}
Johnson H.R., Ake T.B., Ameen M.M., 1993, ApJ 402, 667

\bibitem[]{}
Jorissen A., Mayor M., 1992, A\&A 260, 115

\bibitem[]{}
Jorissen A., Frayer D.T., Johnson H.R., Mayor M., Smith V.V., 1993,
A\&A 271, 463

\bibitem[]{}
Jorissen A., Knapp G.R., 1997, submitted

\bibitem[]{}
Jorissen A., Van Eck S., 1997. In:
Wing R. (ed.) The Carbon Star Phenomenon (IAU
Symp. 177). Kluwer, Dordrecht, in press

\bibitem[]{}
Jorissen A., Van Eck S., Mayor M., Udry S., 1997a,
A\&A, submitted

\bibitem[]{}
Jorissen A., Mowlavi N., Sterken C., Manfroid J., 1997b,
A\&A, in press

\bibitem[]{}
K\"appeler F., Beer H., Wisshak K., 1989, Rep. Prog. Phys. 52, 945

\bibitem[]{}
Keenan P.C., 1950, AJ 55, 172

\bibitem[]{}
Keenan P.C., 1954, ApJ 120, 484

\bibitem[]{} Kholopov P.N., Samus' N.N., Frolov M.S., et al.,
1985, General Catalogue of Variable Stars (4th edition), Moscow, Nauka

\bibitem[]{}
Koornneef J., 1983, A\&A 128, 84

\bibitem[]{}
Lindegren L., 1995, A\&A 304, 61

\bibitem[]{}
Little S.J., Little-Marenin I.R., Hagen-Bauer W., 1987, AJ 94, 981 

\bibitem[]{}
Lloyd Evans T., 1983a, MNRAS 204, 985

\bibitem[]{}
Lloyd Evans T., 1983b, MNRAS 204, 975

\bibitem[]{}
Lloyd Evans T., 1984, MNRAS 208, 447

\bibitem[]{}
Lloyd Evans T., Catchpole R.M., 1989, MNRAS 237, 219

\bibitem[]{}
L\"u P.K., 1991, AJ 101, 2229

\bibitem[]{}
Lundgren K., 1988, A\&A 200, 85

\bibitem[]{}
Lundgren K., 1990, A\&A 233, 21

\bibitem[]{}
Luri X., Arenou F., 1997. In: Perryman M.A.C. (ed.) The Hipparcos and
Tycho Catalogues, ESA-SP 402, in press

\bibitem[]{}
Lutz T.E., 1979, MNRAS 189, 273

\bibitem[]{}
Lutz T.E., Kelker D.H., 1973, PASP 85, 573

\bibitem[]{}
Malmquist K.G., 1936, Stockolms Obs. Medd. No. 26
 
\bibitem[]{}
Mathews G.J., Takahashi K., Ward R.A., Howard W.M., 1986,
ApJ 302, 410

\bibitem[]{}
Mavridis L.N., 1960, PASP 72, 48

\bibitem[]{}
Mayor M., Duquennoy A., Udry S., Andersen J., Nordstr\"om B.,
1996. In: Milone E.F., Mermilliod J.Cl. (eds.) The Origins, Evolution,
and Destinies of Binary Stars in Clusters. ASP Conf. Ser., San
Francisco, p. 190

\bibitem[]{}
McClure R.D., 1984, PASP 96, 117

\bibitem[]{}
McClure R.D., Woodsworth A.W., 1990, ApJ 352, 709

\bibitem[]{}
Mennessier M.-O., Luri X., Figueras F., G\'omez, A.E., Grenier S., Torra J., 1997,
A\&A, in press

\bibitem[]{}
Merrill P.W., 1922, ApJ 56, 457

\bibitem[]{}
Merrill P.W., 1952, ApJ 116, 21

\bibitem[]{}
Miller G.E., Scalo J., 1982, ApJ 263, 259

\bibitem[]{}
Mould J., Aaronson M., 1986, ApJ 303, 10

\bibitem[]{}
Neckel Th., Klare G., 1980, A\&AS 42, 251 

\bibitem[]{}
Neugebauer G., Leighton R.B., 1969. Two-Micron Sky Survey, 
NASA SP-3047 (TMSS)

\bibitem[]{}
Noguchi K., Sun J., Wang G., 1991, PASJ 43, 275

\bibitem[]{}
Pont F., Mayor M., Turon C., VandenBerg D.A., 1997, A\&A, submitted

\bibitem[]{}
Paczy\'nski B., Rudak B., 1980, A\&A 82, 349

\bibitem[]{}
Peery B.F.Jr., 1986. In: New Insights in Astrophysics. ESA-SP 263, p.117

\bibitem[]{}
Price S.D., Murdock T.L., 1983. The Revised AFGL Infrared Sky Survey Catalog
(AFGL-TR-83-0161)

\bibitem[]{}
Rebeirot E., Azzopardi M., Westerlund B.E., 1993, A\&AS 97, 603

\bibitem[]{}
Reid N., Mould J., 1985, ApJ 299, 236

\bibitem[]{}
Richer H.B., 1981, ApJ 243, 744

\bibitem[]{}
Ridgway S.T., Joyce R.R., White N.M., Wing R.F., 1980, ApJ 235, 126 

\bibitem[]{}
Sackmann I.-J., Boothroyd A.I., 1991. In: 
Michaud G., Tutukov A. (eds.) 
Evolution of Stars: The Photospheric Abundance Connection (IAU
Symp. 145). Kluwer, Dordrecht, p. 275

\bibitem[]{}
Scalo J.M., 1976, ApJ 206, 474

\bibitem[]{}
Scalo J.M., Miller G.E., 1981, ApJ 246, 251

\bibitem[]{}
Schaller G., Schaerer G., Meynet G., Maeder A., 1992, A\&AS 96, 269

\bibitem[]{}
Smith H.Jr., 1988, A\&A 198, 365

\bibitem[]{}
Smith V.V., Lambert  D.L.,  1986, ApJ   311,   843

\bibitem[]{}
Smith V.V., Lambert  D.L.,  1988, ApJ   333,   219

\bibitem[]{} 
Smith V.V., Lambert D.L.,  1990, ApJS 72, 387

\bibitem[]{} 
Smith V.V., Plez B., Lambert D.L., Lubowich D.A., 1995, ApJ 441, 735

\bibitem[]{} 
Smith V.V., Cunha K., Jorissen A., Boffin H.M.J., 1996,
A\&A 315, 179

\bibitem[]{} 
Smith V.V., Cunha K., Jorissen A., Boffin H.M.J., 1997, A\&A, in press

\bibitem[]{}
Stein R.W., 1991, ApJ 377, 669

\bibitem[]{}
Stephenson C.B., 1984. The General Catalogue of Galactic S Stars, Publ. Warner \&
Swasey Observ. 3, 1
 
\bibitem[]{}
Takayanagi W., 1960, PASJ 12, 314

\bibitem[]{}
Turon C. et al., 1992a. The HIPPARCOS Input Catalogue, ESA SP-1136

\bibitem[]{}
Turon C. et al., 1992b, Bull. Inform. CDS 41, 9

\bibitem[]{}
Turon C. et al., 1992c, Bull. Inform. CDS 43, 5

\bibitem[]{}
van der Veen W.E.C.J., Habing H.J., 1988, A\&A 194, 125

\bibitem[]{}
van Leeuwen F., Feast M.W., Whitelock P.A., Yudin B., 1997, MNRAS, in press

\bibitem[]{}
Vanture A.D., Wallerstein G., Brown J.A., 1994, PASP 106, 835

\bibitem[]{}
Warner B., 1965, MNRAS 129, 263

\bibitem[]{}
Wagenhuber J., Tuchman Y., 1996, A\&A 311, 509

\bibitem[]{}
Westerlund B.E., Olander N., Hedin B., 1981, A\&AS 43, 267

\bibitem[]{}
Westerlund B.E., Azzopardi M., Breysacher J., Rebeirot E., 1991, A\&AS
91, 425 (with erratum in A\&AS 92, 683)

\bibitem[]{}
Westerlund B.E., Azzopardi M., Breysacher J., Rebeirot E., 1995, A\&A 303, 107

\bibitem[]{}
Willems F.S., de Jong T., 1986, ApJ 309, L39
 
\bibitem[]{}
Wing R. F., Yorka, S. B., 1977, MNRAS 178, 383

\bibitem[]{}
Wood P.R., Bessell M.S., Fox M.W., 1983, ApJ 272, 99

\bibitem[]{}
Wu C.-C., Ake T.B., Boggess A., Bohlin R.C., Imhoff C.L., Holm A.V., Levay Z.G.,
Panek R.J., Chiffer F.H.III, Turnrose B.E., 1983. The Ultraviolet Spectral Atlas,
NASA IUE Newsletter 22 

\bibitem[]{}
Yorka S.B., Wing R.F., 1979, AJ 84, 1010

\end{thebibliography}
\end{document}